\documentclass[epjCONF]{svjour}
\usepackage{graphics}
\usepackage{graphicx}
\usepackage[varg]{txfonts} 
\usepackage[latin1]{inputenc}
\usepackage{hyperref}

\session-title{%
19$^{\textnormal{\footnotesize th}}$ International %
IUPAP Conference on Few-Body Problems in Physics%
}

\begin{document}
\title{Toward the Application of Three-Dimensional Approach to Few-body Atomic Bound States}%
\author{%
M. R. Hadizadeh\thanks{\email{hadizade@ift.unesp.br}}
\and %
L. Tomio\thanks{\email{tomio@ift.unesp.br}}
}
\institute{Instituto de F\'{\i}sica Te\'{o}rica (IFT), Universidade Estadual Paulista (UNESP), Barra Funda,
01140-070, S\~{a}o Paulo, Brazil. }
\abstract{ The first step toward the application of an effective non partial wave (PW) numerical approach to few-body atomic bound states has been taken. The two-body transition amplitude which appears in the kernel of three-dimensional Faddeev-Yakubovsky integral equations is calculated as function of two-body Jacobi momentum vectors, i.e.
as a function of the magnitude of initial and final momentum vectors and the angle between them.
For numerical calculation the realistic interatomic interactions HFDHE2, HFD-B, LM2M2 and TTY are used.
The angular and momentum dependence of the fully off-shell transition amplitude is studied at negative energies.
It has been numerically shown that, similar to the nuclear case, the transition amplitude exhibits a characteristic
angular behavior in the vicinity of $^4$He dimer pole.
} 
\maketitle
%
%
%
\section*{Introduction}

In recent years the $^4$He trimer and tetramer have been the center
of several theoretical investigations (see, for example, Refs.
\cite{Nielsen-JPB3,Roudnev-CPL328,Kolganova-JPB31} and references
therein). From all employed methods in these studies, the
Faddeev-Yakubovsky (FY) schemas are perhaps most attractive
since they reduces the Schr\"odinger equation for three (four)
particle systems into a coupled set (two coupled sets) of integral or
differential equations which can be used to study the bound and
scattering states in a rigorous way. The differential form of FY equations has been successfully applied in nuclear bound states calculations, but there are limitations in its application to atomic systems. The limitation arises from eccentricities of the interatomic interactions, since interatomic interactions often
contain very strong short range repulsion which leads to tedious and cumbersome numerical procedure. In calculations of atomic systems, because of the short range correlations, one needs a large number of PWs to obtain the converged results. To overcome this problem few numerical techniques are developed, the tensor-trick method \cite{Schellingerhout-PRA40,Schellingerhout-FBS11}, representation of Faddeev equations in Cartesian coordinate \cite{Carbonell-FBS15}, the operator form of Faddeev equations in total angular
momentum representation \cite{Kostrykin-FBS6} and also a hybrid method \cite{Roudnev-FBS37}.
The three and four-body atomic bound
states have been also studied with short-range forces and large
scattering length at leading order in an Effective field theory
approach \cite{Platter-PRA70}-\cite{Platter-PLB607}, but these
investigations are also based on PW decomposition and the
interactions are restricted to only s-wave sector.

By these considerations we are going to extend a numerical method, which has been successfully applied to nuclear bound and scattering systems and avoids the PW representation and its complexity, to atomic bound states.
It should be clear that the building blocks to the few-body calculations without angular momentum
decomposition are two-body off-shell transition amplitudes, which depend on the magnitudes of the initial and
final Jacobi momenta and the angle between them. Elster \emph{et al.} have calculated the $NN$ transition amplitude for spinless particles in a non PW representation by using the Malfliet-Tjon type potentials \cite{Elster-FBS24}. Our aim in this paper is to calculate the matrix elements of the fully off-shell two-body transition amplitude at negative energies for realistic interatomic interactions, we study the momentum and angle dependence of transition amplitudes.
This paper is organized as follows. In section \ref{sec.potentials} we represent the explicit form of studied interatomic interactions in configuration and momentum spaces. In sections \ref{sec.dimer} and \ref{sec.t-matrix} we present our numerical results for homogenous and inhomogenous Lippmann-Schwinger equations in a non PW representation. An outlook is provided in section \ref{sec.outlook}.

\section{$^4$He-$^4$He Interatomic Interactions} \label{sec.potentials}

\begin{figure*}[hbt]
\centering
\includegraphics*[width=14.0cm]{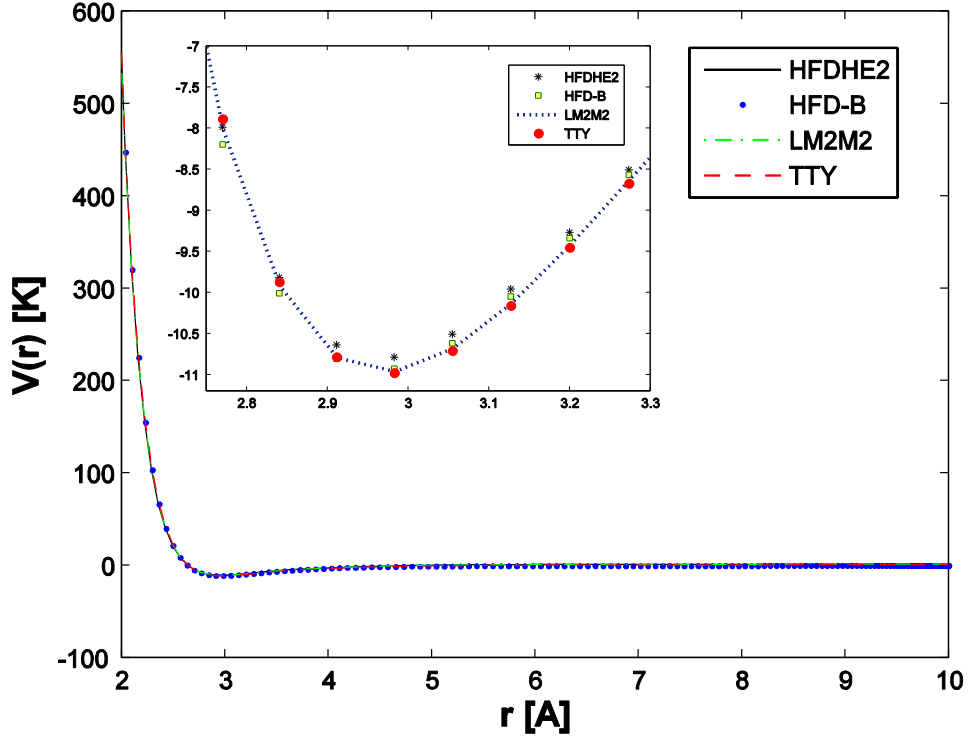}
\caption{The interatomic $^4$He-$^4$He potentials as a function of the distance $r$ between the atoms. The region around minima of the potentials is shown in the inset of the figure.}
\label{fig.potentials-configuration}
\end{figure*}

In this study as $^4$He-$^4$He interatomic interactions we use the realistic HFDHE2~\cite{Aziz_JCP79}, HFD-B~\cite{Aziz_MP61}, LM2M2~\cite{Aziz_JCP94} and TTY~\cite{Tang_PRL74} potentials.
The semi-empirical HFDHE2, HFD-B and LM2M2 potentials, which are constructed by Aziz and collaborators, have the general form
\begin{eqnarray}
\label{HHL}
  V(r)=\varepsilon \, \biggl (V_a(x)+V_b(x)\biggr ),
\end{eqnarray}
where $x=\frac{r}{r_m}$, $r$ and $r_m$ are expressed in the unit {\AA}.
The terms $V_a(x)$ and $V_b(x)$ read
\begin{eqnarray}
V_a(x)= \Biggl \{ \begin{array}{cl}
  A_a \left (\sin \Bigl [\displaystyle \frac{2\pi(x-x_1)}{x_2-x_1}-
\frac{\pi}{2} \Bigr ]+1 \right ), \quad & x_1 \leq x \leq x_2    \\ \\
0, & x\not\in[x_1,x_2]\,.
\end{array}
\end{eqnarray}
\begin{eqnarray}
V_b(x)=A \, e^{ \left(-\alpha x  +\beta x^2 \right)} -\biggr ( \frac{C_6}{x^6} + \frac{C_8}{x^8} + \frac{C_{10}}{x^{10} } \biggr ) F(x)\,,
\end{eqnarray}
and the function $F(x)$ is given by
\begin{eqnarray}
F(x)= \Biggl \{ \begin{array}{cr}
 e^{ -\left( \frac{D}{x} -1 \right)^2 },	&   x\leq D    \\ \\
1,  &   x > D .
\end{array}
\end{eqnarray}
The parameters of the HFDHE2, HFD-B and LM2M2 potentials are
given in Table \ref{table.Aziz}. The explicit form of the theoretical TTY potential is
\begin{eqnarray}
 V(r)=A\, \bigl( V_{\rm ex}(r)+V_{\rm disp}(r) \bigr),
\end{eqnarray}
where $r$ stands for the distance between the $^4$He atoms given in
atomic length units. The function $V_{\rm ex}(r)$ has the form
\begin{eqnarray}
V_{\rm ex}(r)=D\,r^p\, e^{(-2\beta r)},
\end{eqnarray}
with $p=\displaystyle\frac{7}{2\beta}-1\,$. The function
$V_{\rm disp}(r)$ reads
\begin{eqnarray}
  V_{\rm disp}(r)=-\sum\limits_{n=3}^N C_{2n}\,f_{2n}(r)\,r^{-2n},
\end{eqnarray}
the coefficients $C_{2n}$ are calculated via the recurrency relation
\begin{eqnarray}
C_{2n}=\left(\displaystyle\frac{C_{2n-2}}{C_{2n-4}}\right)^3\,C_{2n-6},
\end{eqnarray}
and the functions $f_{2n}(r)$ are given by
\begin{eqnarray}
f_{2n}(r)=1-e^{ \bigl(-b(r) \,r \bigr)}\,\sum\limits_{k=0}^{2n} \displaystyle\frac{\bigl(b(r)\, r \bigr)^k}{k!},
\end{eqnarray}
where
\begin{eqnarray}
b(r)=2\beta-\left[\displaystyle\frac{7}{2\beta}-1\right]\,
\displaystyle\frac{1}{r}\,.
\end{eqnarray}
The parameters of the TTY potential are given in Table \ref{table.TTY}. In Figure (\ref{fig.potentials-configuration}) all potentials are plotted as function of distance between the $^4$He atoms. Moreover, in the inset to this figure the region around the minima of the potentials is shown for providing a better comparison.

\begin{figure*}[hbt]
\centering
\begin{tabular}{ccc}
\includegraphics*[width=5.2cm]{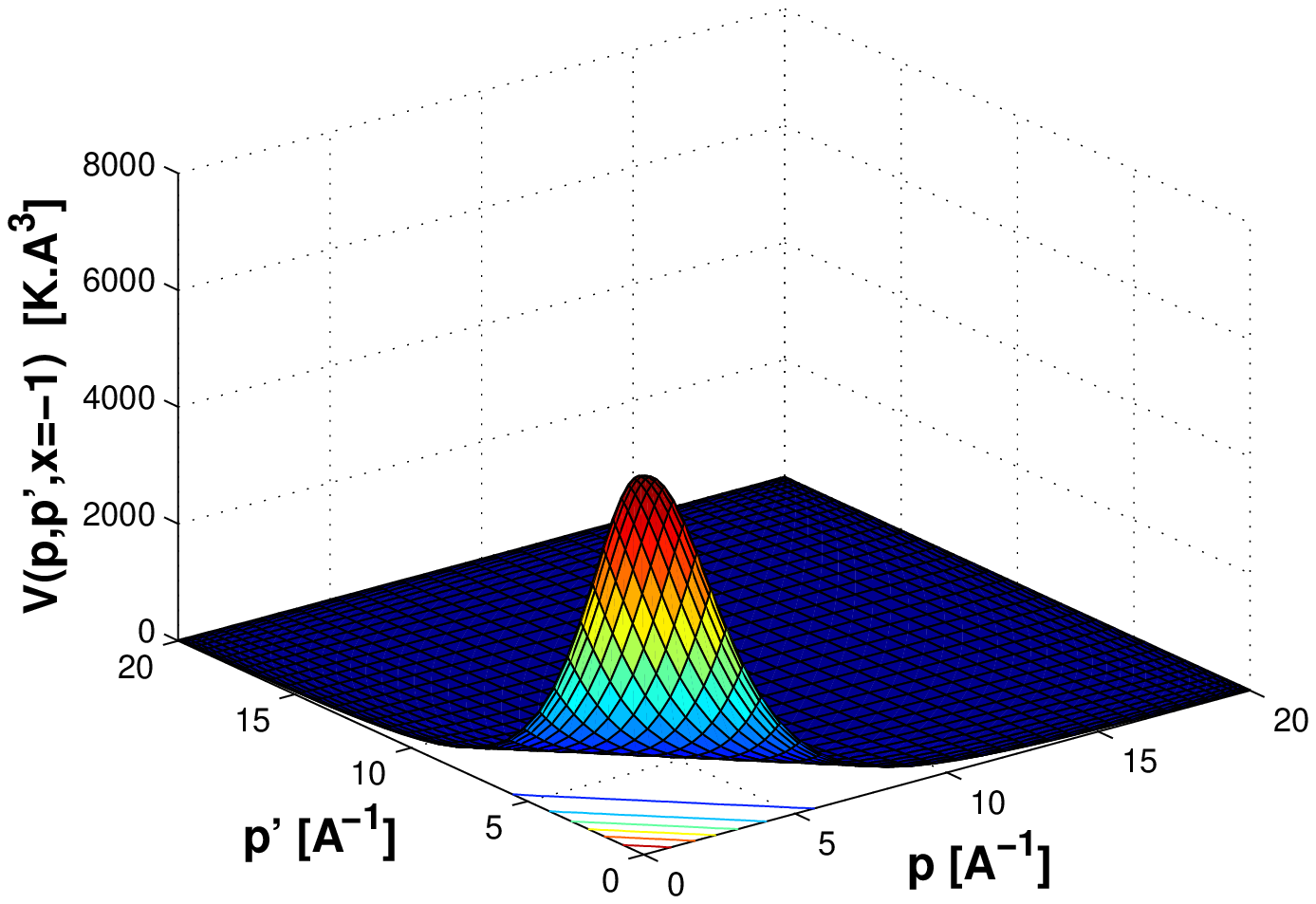} &
\includegraphics*[width=5.2cm]{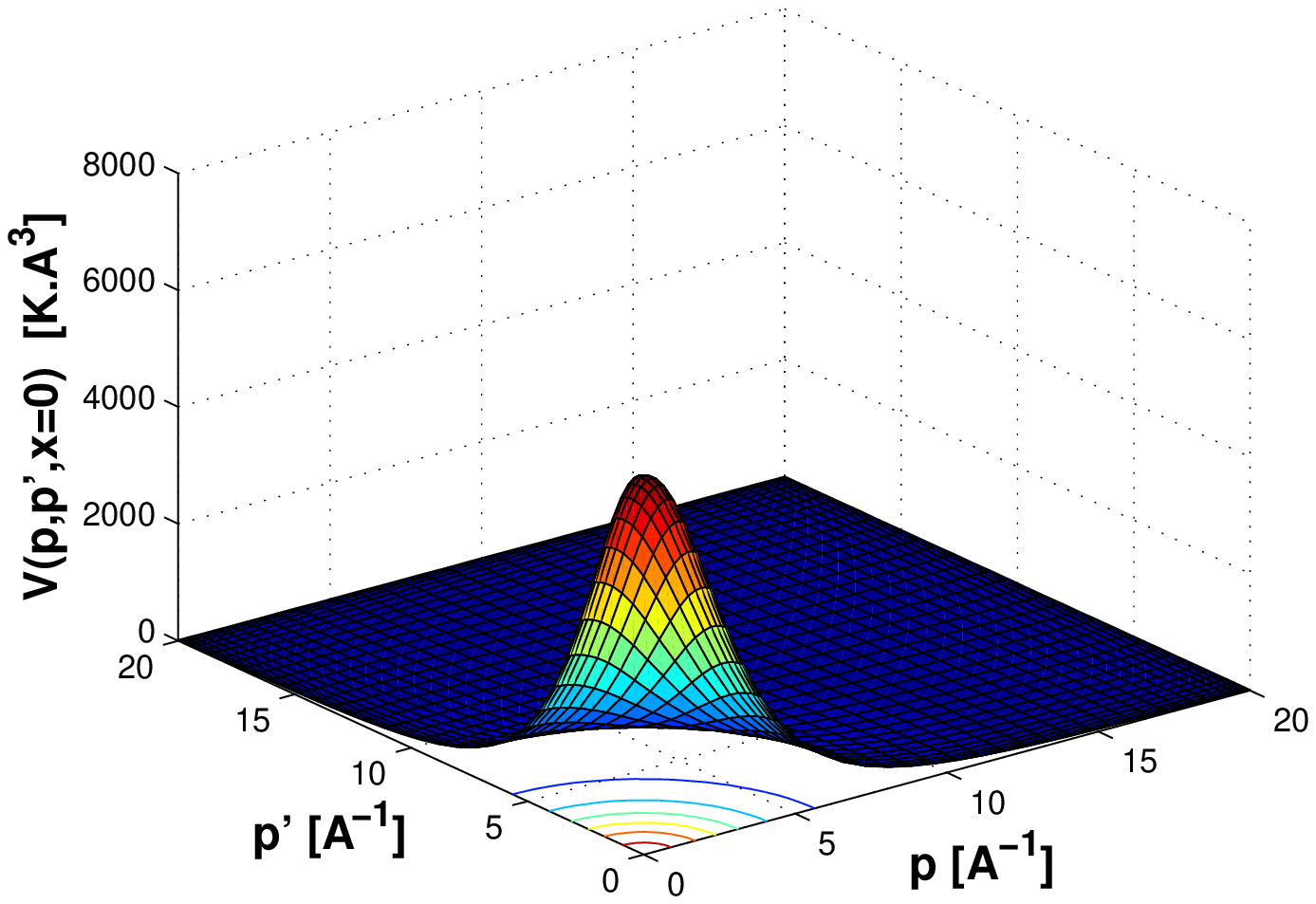} &
\includegraphics*[width=5.2cm]{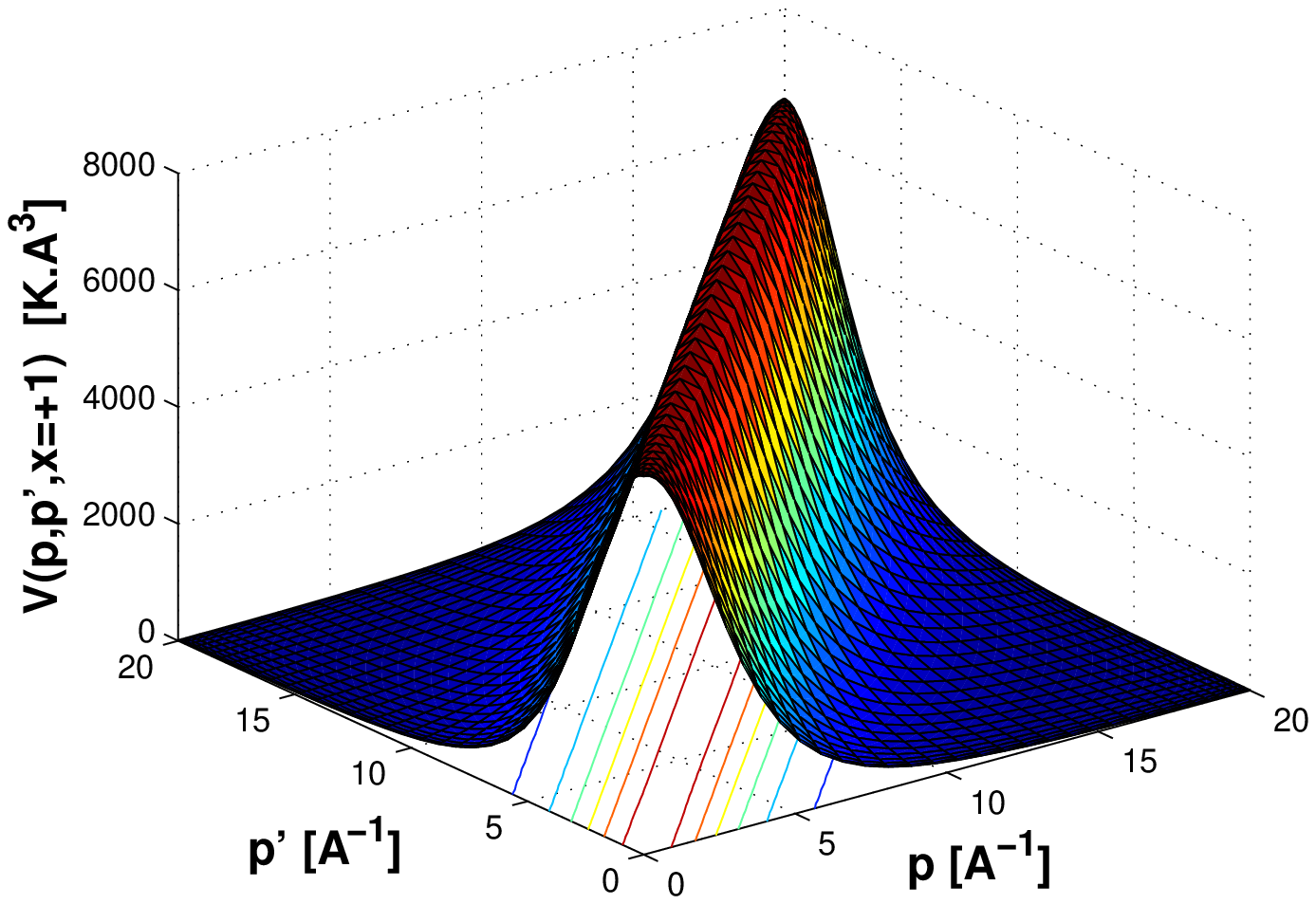} \\
HFDHE2 & HFDHE2 & HFDHE2 \\
\includegraphics*[width=5.2cm]{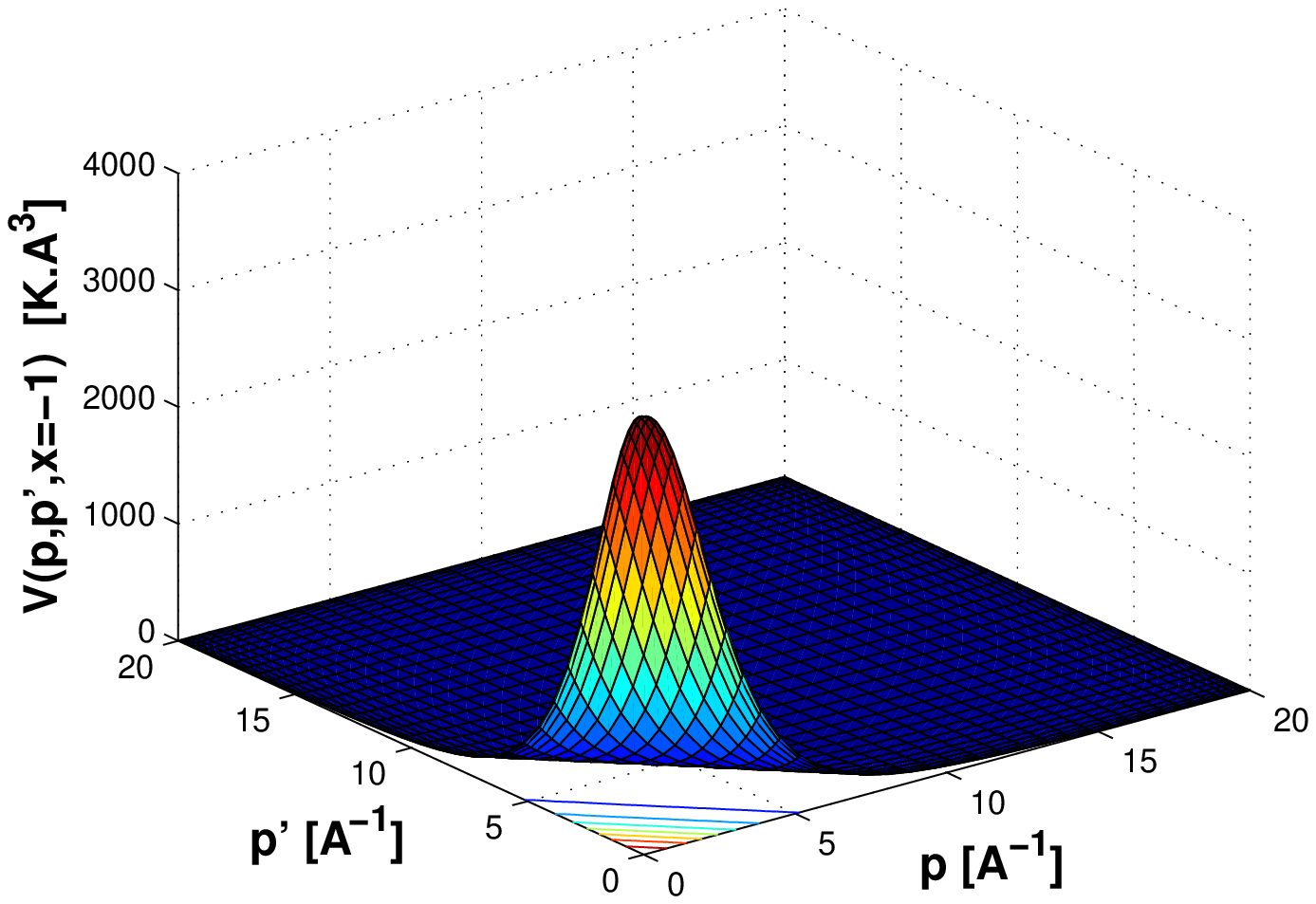} &
\includegraphics*[width=5.2cm]{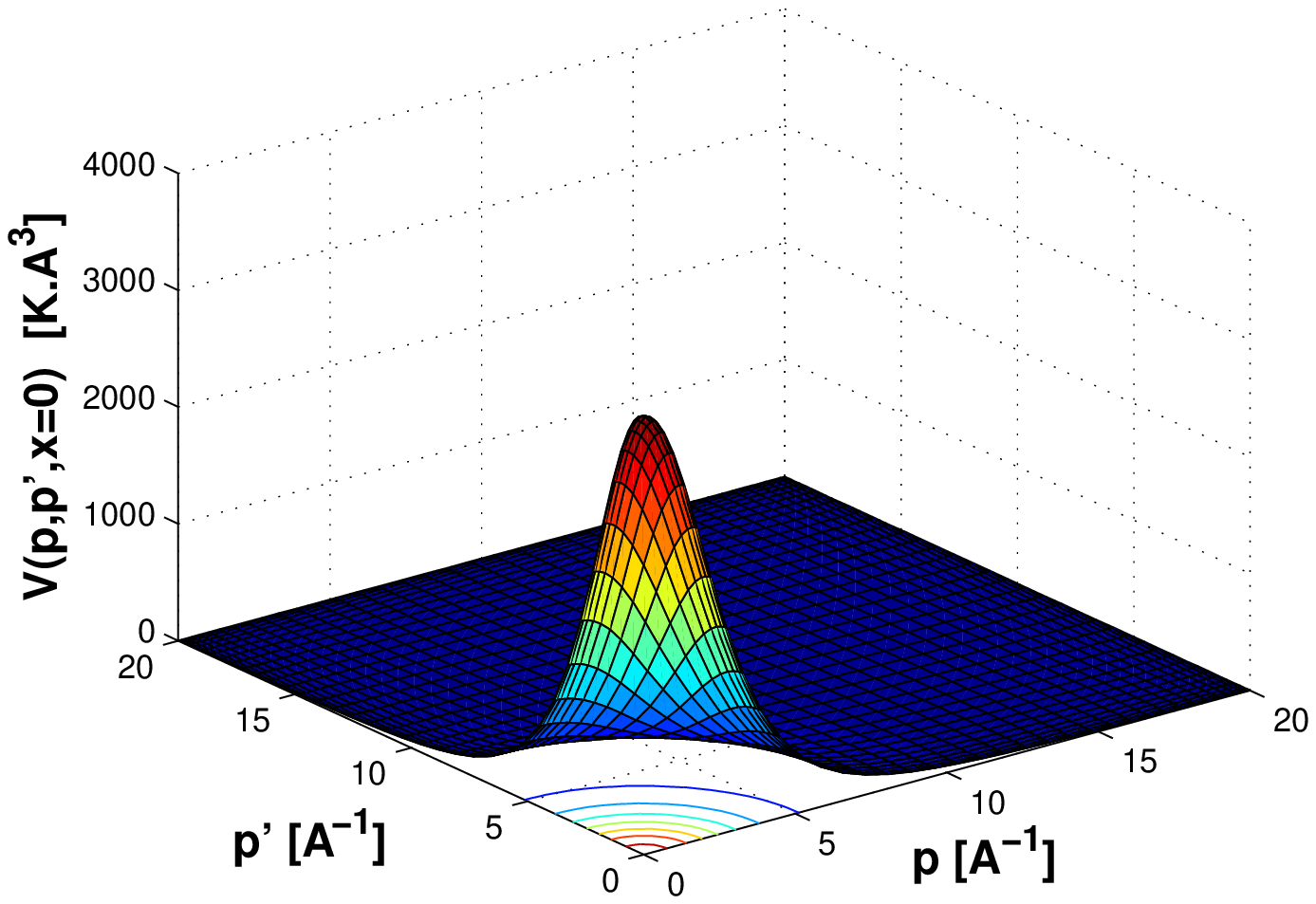} &
\includegraphics*[width=5.2cm]{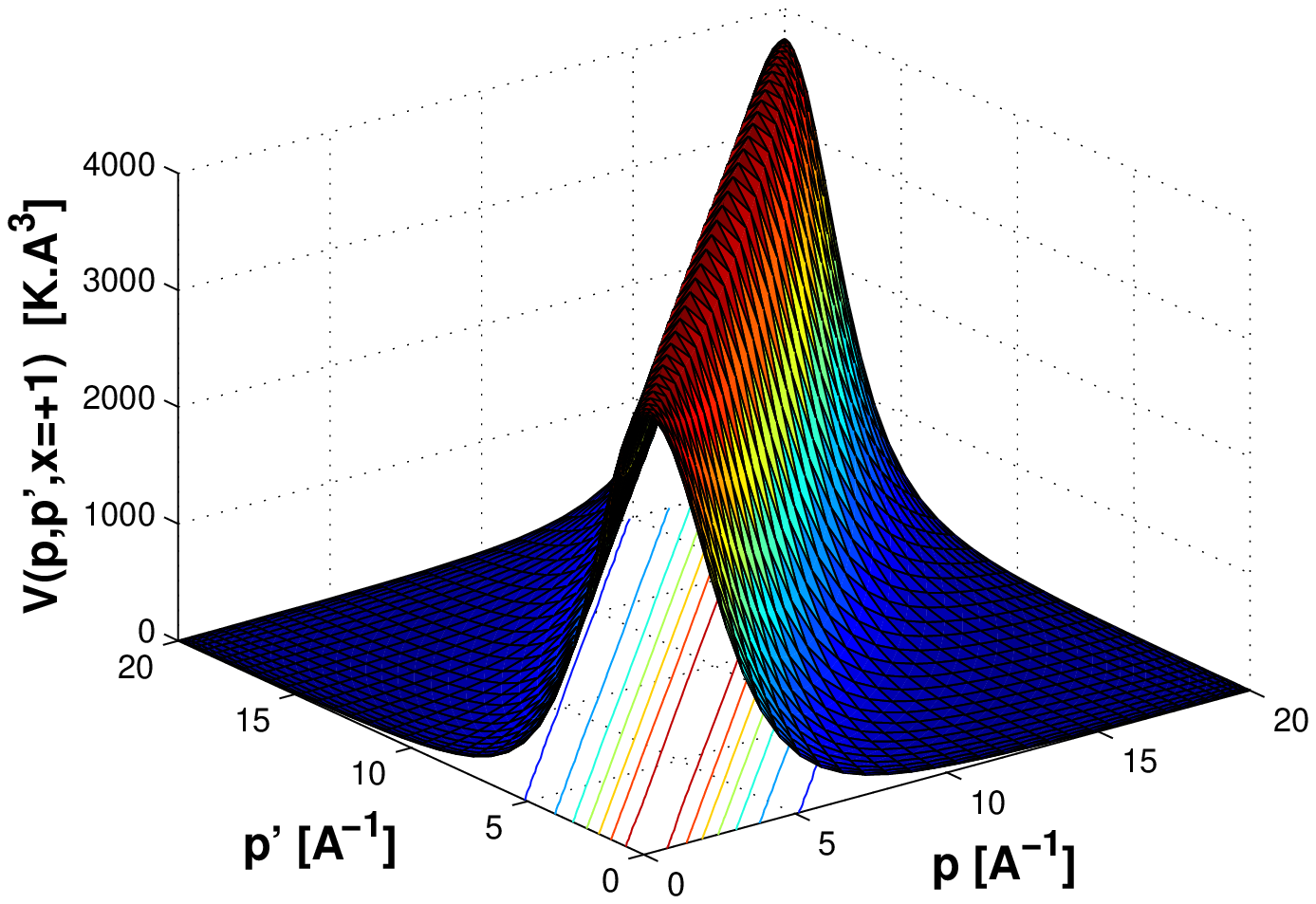} \\
HFD-B & HFD-B & HFD-B \\
\includegraphics*[width=5.2cm]{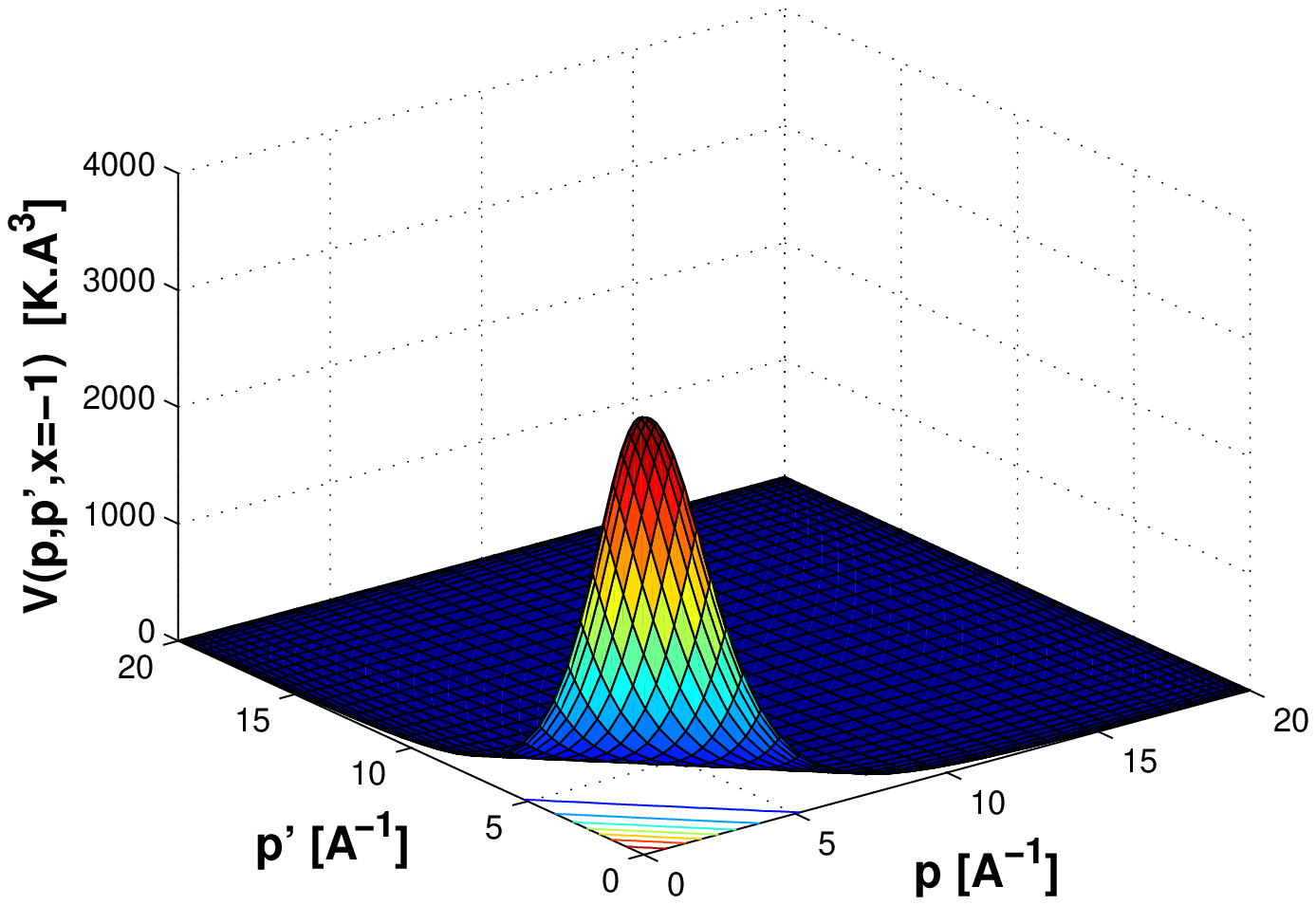} &
\includegraphics*[width=5.2cm]{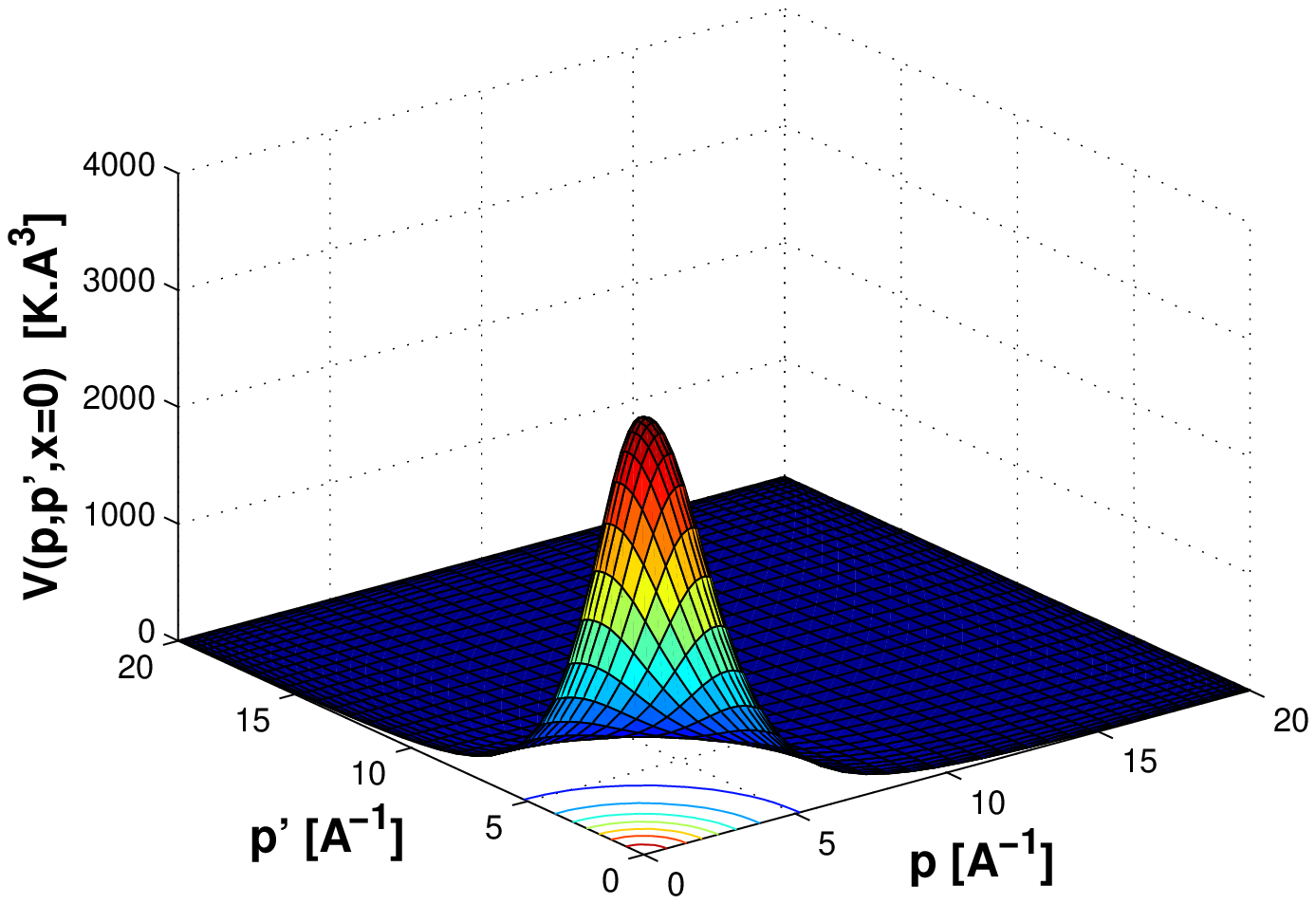} &
\includegraphics*[width=5.2cm]{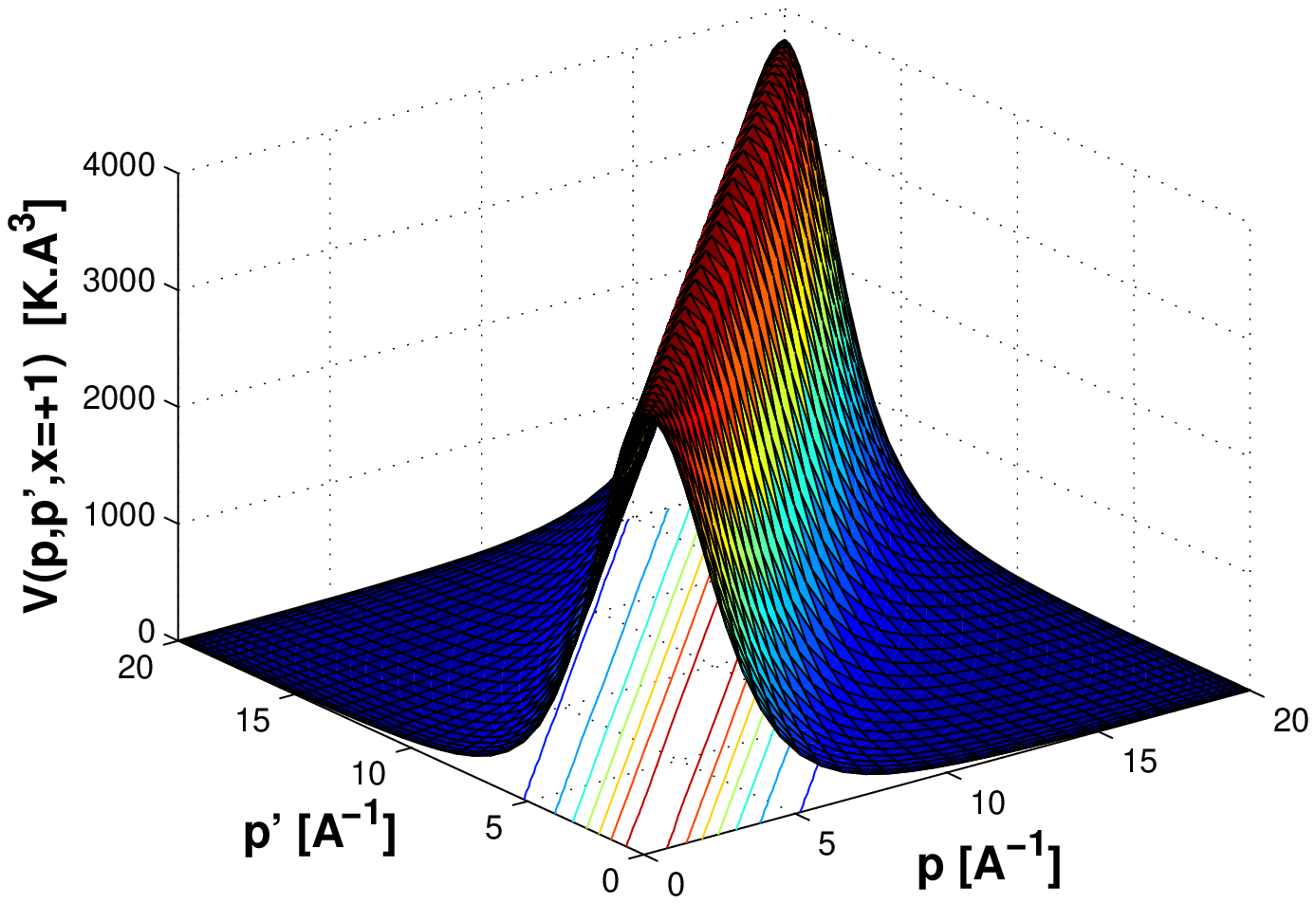} \\
LM2M2 & LM2M2 & LM2M2 \\
\includegraphics*[width=5.2cm]{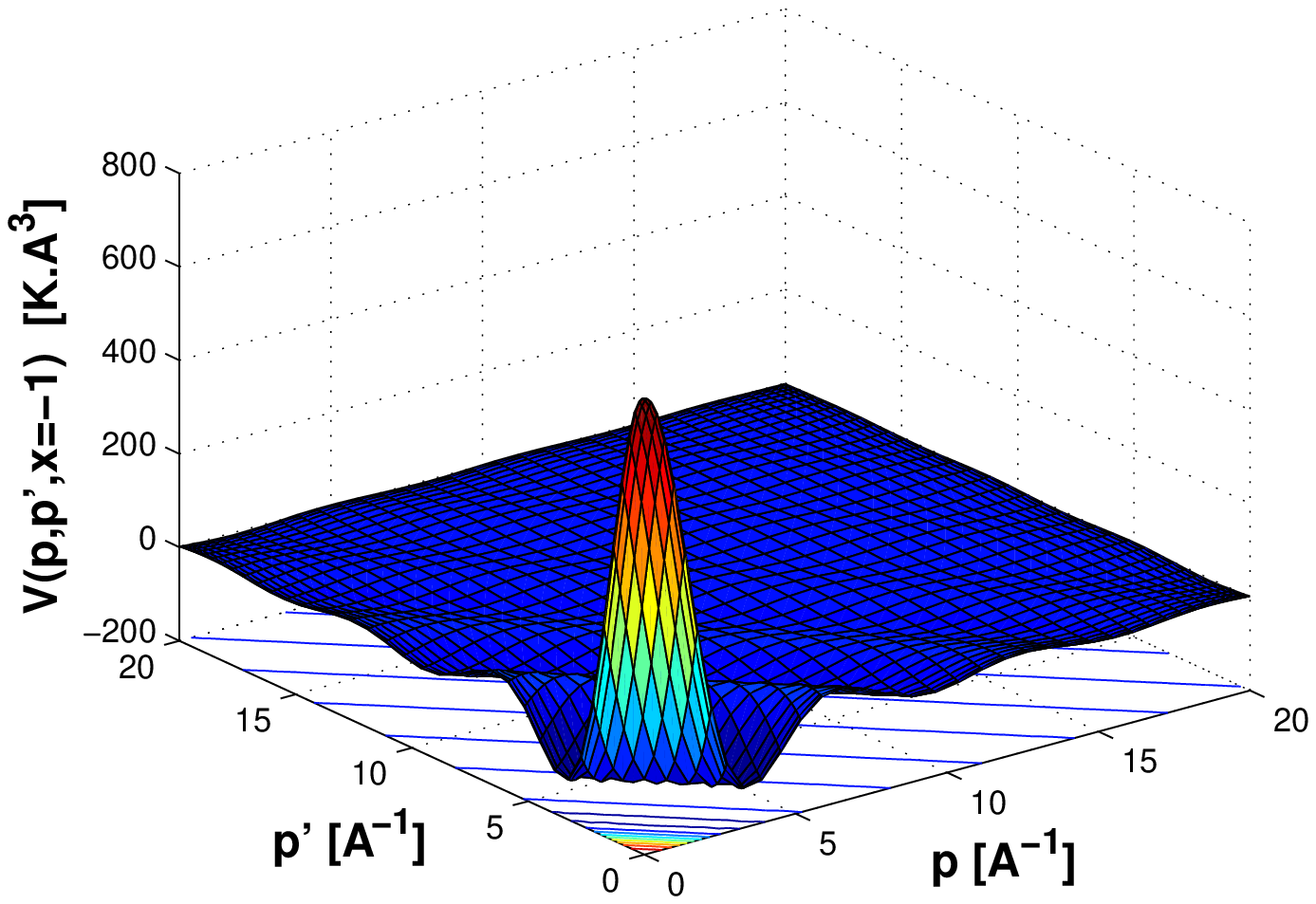} &
\includegraphics*[width=5.2cm]{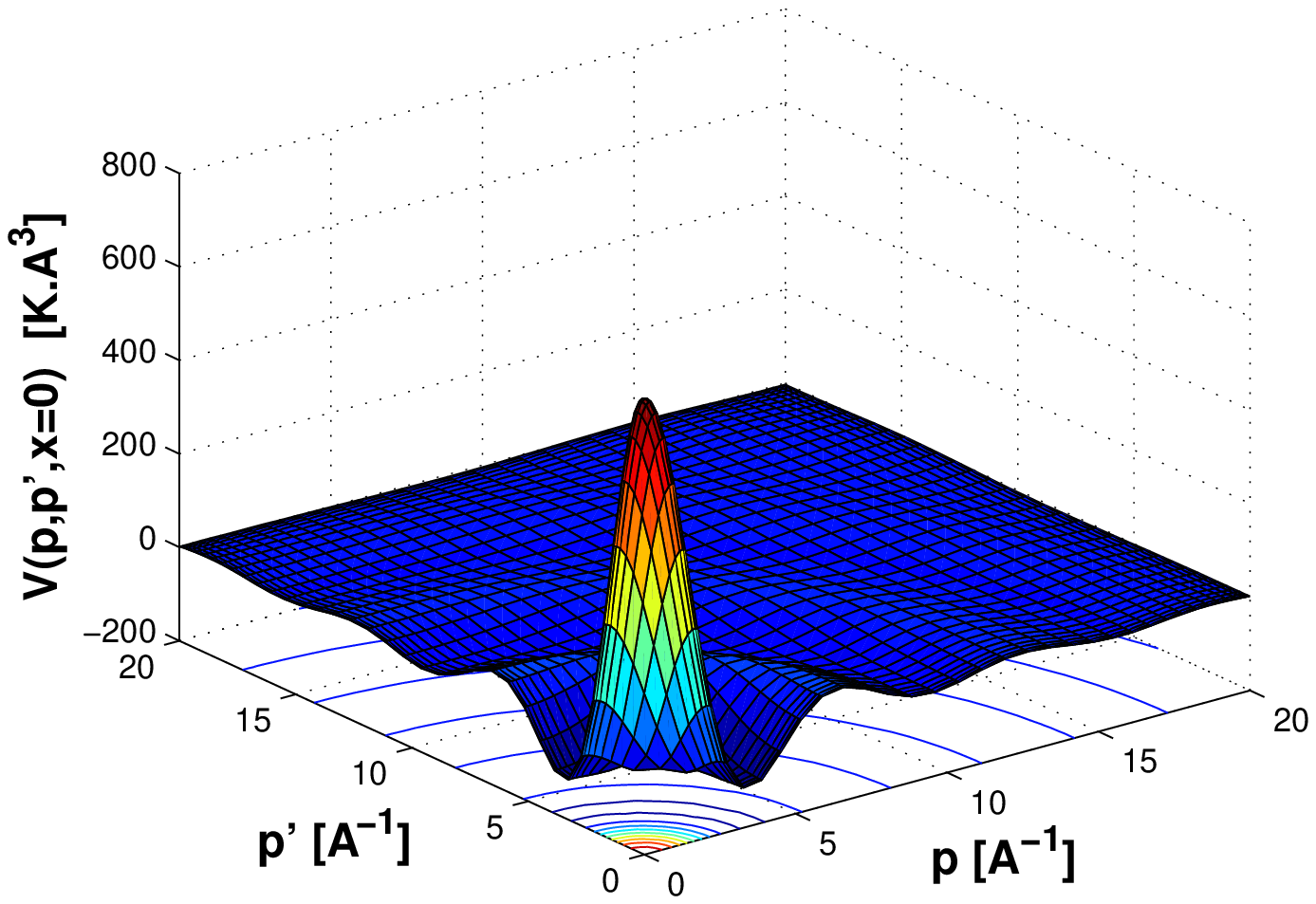} &
\includegraphics*[width=5.2cm]{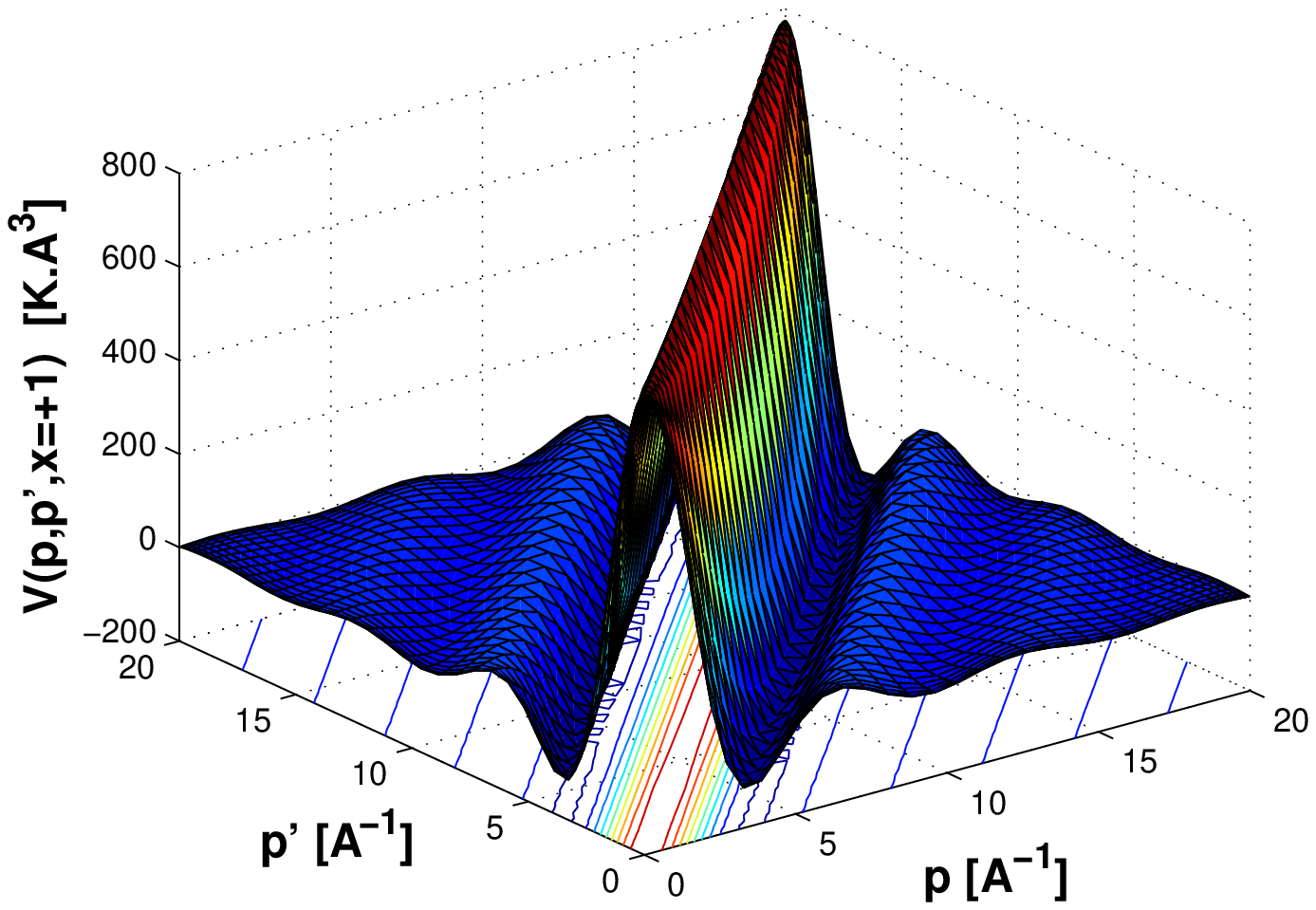} \\
TTY & TTY & TTY \\
\end{tabular}
\caption{Momentum-space representation of interatomic $^4$He-$^4$He potentials $V(p,p',x_{pp'})$ in fixed angles $x_{pp'}=0,\pm1$.} \label{fig.potentials-momentum}
\end{figure*}

\begin{table}
\caption{The parameters of the $^4$He$-$$^4$He interactions which are constructed by Aziz and collaborators.}
\label{table.Aziz}
\begin{tabular}{llll}
\hline\noalign{\smallskip}
Parameter & HFDHE2  & HFD-B  & LM2M2  \\\hline
   $\varepsilon$ [K]      &    10.8       &  10.948    &   10.97 \\
   $ r_m $ [\AA]          &   2.9673      &  2.963     &   2.9695  \\
   $A$                    &   544850.4    &  184431.01 &  189635.353 \\
   $\alpha$               & 13.353384     & 10.43329537 & 10.70203539\\
   $\beta$                &     0         & $-2.27965105$& -1.90740649 \\
   $C_6$                  & 1.3732412     & 1.36745214 & 1.34687065  \\
   $C_8$                  & 0.4253785     & 0.42123807 & 0.41308398 \\
   $C_{10}$               & 0.178100      & 0.17473318 & 0.17060159 \\
   $D$                    & 1.241314      & 1.4826     & 1.4088 \\
   $A_a$                  & $-$           & $-$        & 0.0026  \\
   $x_1$              & $-$           & $-$        & 1.003535949\\
   $x_2$              & $-$           & $-$        & 1.454790369\\
\noalign{\smallskip}\hline
\end{tabular}
\end{table}

\begin{table}
\caption{The parameters of the $^4$He$-$$^4$He TTY potential.}
\label{table.TTY}
\begin{tabular}{lllllllll}
\hline\noalign{\smallskip}
 &&&  $A$ [K]    &&&&&  $315766.2067$                 \\
 &&& $\beta$ $\bigl[$(a.u.)$^{-1}$$\bigr]$ &&&&& $1.3443$  \\
 &&&   $D$        &&&&&   $7.449$  \\
 &&&  $N$        &&&&&   $12$     \\
 &&&  $C_6$      &&&&&   $1.461$  \\
 &&&  $C_8$      &&&&&   $14.11$  \\
 &&&  $C_{10}$   &&&&&   $183.5$  \\
\noalign{\smallskip}\hline
\end{tabular}
\end{table}

In order to be able to implement the introduced interatomic interactions in few-body atomic bound and scattering state calculations in momentum space, we need to transform these potentials to momentum space. The matrix elements of the potentials can be obtained by following relation
\begin{eqnarray} \label{eq.pot-moment}
V({\bf p},{\bf p'}) &\equiv& V(p,p',x_{pp'}) \nonumber \\ &=& \frac{1}{2 \pi^2 q} \nonumber \int_0^{\infty} dr \, r \sin(qr) \, V(r) \quad;
q=|{\bf p}-{\bf p'}|\neq 0 \nonumber \\ \nonumber \\
&=& \frac{1}{2 \pi^2} \int_0^{\infty} dr \, r^2 \, V(r) \quad\quad\quad\quad \,\,;  q=0
\end{eqnarray}
where $p$ and $p'$ are magnitudes of initial and final two-body Jacobi momentum vectors, $x_{pp'}$ is the angle between them and $q=(p^2+p'^2-2pp'x_{pp'})^\frac{1}{2}$ is the difference between them. Clearly the knowledge of the structure and range of potentials is important in the calculation of two-body transition amplitudes, which appear in the kernel of the integral equations of two-, three- and four-body bound and scattering calculations. In Figure (\ref{fig.potentials-momentum}) the momentum dependence of the potentials are shown at fixed angles $x_{pp'}=0,\pm1$. As shown all potentials have similar behavior. The ridge around $p=p'$ arises from strong repulsive core. The behavior of the potentials at forward angle, i.e. $x_{pp'}=+1$, is different from other angles, and according to Eq. (\ref{eq.pot-moment}) the value of the potentials in this angle is fixed for $p=p'$. Note that the potentials vanish at enough large values of Jacobi momenta. i.e. $p^{max}=20 \, \AA^{-1}$. We should mention that for calculation of matrix elements of potential, Eq. (\ref{eq.pot-moment}), the cutoff value of $r^{max}=30 \, \AA$ and 200 mesh points have been used.

\section{$^4$He Dimer}\label{sec.dimer}
The $^4$He dimer can be described by homogeneous Lippmann-Schwinger integral equation:
\begin{eqnarray} \label{eq.dimer}
\psi_d ({\bf p})=\frac{1}{E_d-\frac{p^2}{m}}\int d^3p' \, V({\bf p},{\bf p'}) \, \psi_d ({\bf p'}).
\end{eqnarray}
This integral equation can be solved numerically by direct or iterative methods. We have solved this integral equation by direct method and the numerical results for dimer binding energy by using the introduced interatomic interactions are given in Table (\ref{table:dimer}) in comparison to corresponding PW and experimental results. In our numerical calculations the $^4$He atom mass is defined by $\frac{\hbar^{2}}{m}=12.12 \,{\mathrm{K \, \AA^{2}}}$.
For discretization of the continuous momentum and angle variables we have used the quadrature Gauss-Legendre by using linear mapping for all variables. The number of mesh grids for Jacobi momenta and angle variable are 200 and 150 correspondingly.

\begin{table}[hbt]
\caption {The calculated dimer binding energy in unit of $mK$ for realistic interatomic potentials in comparison to corresponding PW and experimental data.}
\begin{tabular}{ccccccccccccccccccccccccccccccccccccccccccccccccc}
\hline\noalign{\smallskip}
          &   HFDHE2   & HFD-B    & LM2M2     & TTY\\ \hline
   PW \cite{Kolganova-PPN40} &   -0.83012 & -1.68541 &  -1.30348 &  -1.30962  \\
   Presnet     &   -0.83011 & -1.68540 & -1.30347  & -1.30962   \\ \hline
   EXP. \cite{Grisenti-PRL85}   && -1.1$_{-0.2}^{+0.3}$ &\\
\noalign{\smallskip}\hline
\end{tabular}
\label{table:dimer}
\end{table}

\section{Two-Body Transition Amplitude}\label{sec.t-matrix}
The building blocks for few-body bound and scattering state calculations are two-body transition amplitudes $T$ which follow the inhomogenous Lippmann-Schwinger equation
\begin{eqnarray}
 T = V + V G_0 T, \label{eq.T-operator}
\end{eqnarray}
where $V$ is the two-body, e.g. two-atom, potential and $G_0=(E-H_0)^{-1}$
is free two-body propagator. For momentum space calculations one needs the matrix elements of transition amplitude in desired energy $E$ which can be obtained by representation of Eq. (\ref{eq.T-operator}) in two-body basis states \cite{Elster-FBS24}
\begin{eqnarray}\label{eq.T-momentum}
T({\bf p'},{\bf p};E)=V({\bf p'},{\bf p}) + \int d^3 p''
\frac {V({\bf p'},{\bf p''})}{E-\frac {p''^2}{m}}
T({\bf p''},{\bf p};E). \nonumber \\
\end{eqnarray}

In order to solve this three-dimensional integral equation directly without
employing PW projection, we have to define a suitable coordinate
system. To this aim we choose vector ${\bf p}$ parallel to $z-$axis and vector ${\bf p'}$
in the $x-z$ plane and express the integration vector ${\bf p''}$ with respect to them. By this considerations Eq. (\ref{eq.T-momentum}) can be written explicitly as
\begin{eqnarray} \label{eq.T-coordinate}
T(p',p,x_{p\,p'};E) &=& V(p',p,x_{p\,p'}) \nonumber \\ && \hspace{-2.7cm} + \int_0^{\infty} dp'' p''^2 \int_{-1}^{1} dx_{p\,p''}
\int_0^{2\pi} d\varphi''  \frac{1}{E-\frac{p''^2}{m}} \, V(p',p'',x_{p'\,p''}) \nonumber \\ && \hspace{-2.3cm} \times \,
T(p'',p,x_{p\,p''};E),
\end{eqnarray}
where
\begin{eqnarray} \label{eq.T-variables}
x_{p\,p'} &=& {\bf \hat p} \cdot {\bf \hat p}', \nonumber \\
x_{p\,p''}&=& {\bf \hat p} \cdot {\bf \hat p}'', \nonumber \\
x_{p'\,p''} &=& {\bf \hat p}' \cdot {\bf \hat p}'' =x_{p\,p'}x_{p\,p''} +\sqrt{1-x_{p\,p'}^2} \sqrt{1-x_{p\,p''}^2} \cos{\varphi''}. \nonumber \\
\end{eqnarray}

The $\varphi''$ integration acts only on $V(p',p'',x_{p'\,p''})$, so this integration can be carried out separately as
\begin{eqnarray} \label{eq.V-phi-integration}
v(p',p'',x_{p\,p'},x_{p\,p''})\equiv \int_0^{2\pi} d\varphi'' V(p',p'',x_{p'\,p''}),  \label{eq.phi-integration}
\end{eqnarray}
and consequently the integral equation (\ref{eq.T-coordinate}) can be written as
\begin{eqnarray}  \label{eq.T-final}
T(p',p,x_{p\,p'};E) &=& \frac{1}{2\pi} v(p',p,x_{p\,p'},1) \nonumber \\ && \hspace{-1.cm} + \int_0^{\infty} dp'' p''^2
\int_{-1}^{1} dx_{p\,p''}  \frac{v(p',p'',x_{p\,p'},x_{p\,p''})}{E-\frac{p''^2}{m}}
\nonumber \\ && \hspace{-0.5cm}  \times   T(p'',p,x_{p\,p''};E).
\end{eqnarray}

\begin{figure*}[hbt]
\centering
\begin{tabular}{ccc}
\includegraphics*[width=5.2cm]{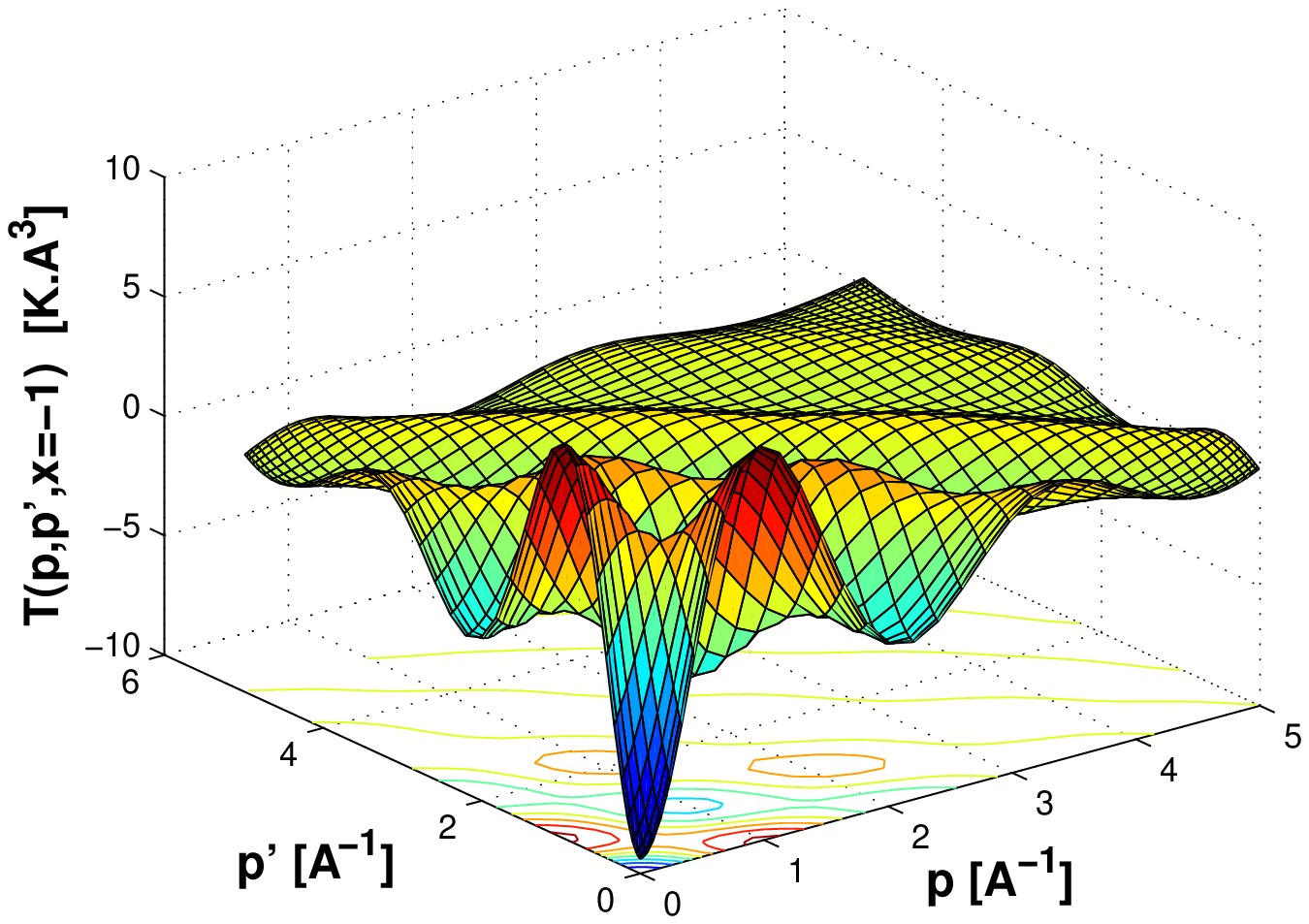} &
\includegraphics*[width=5.2cm]{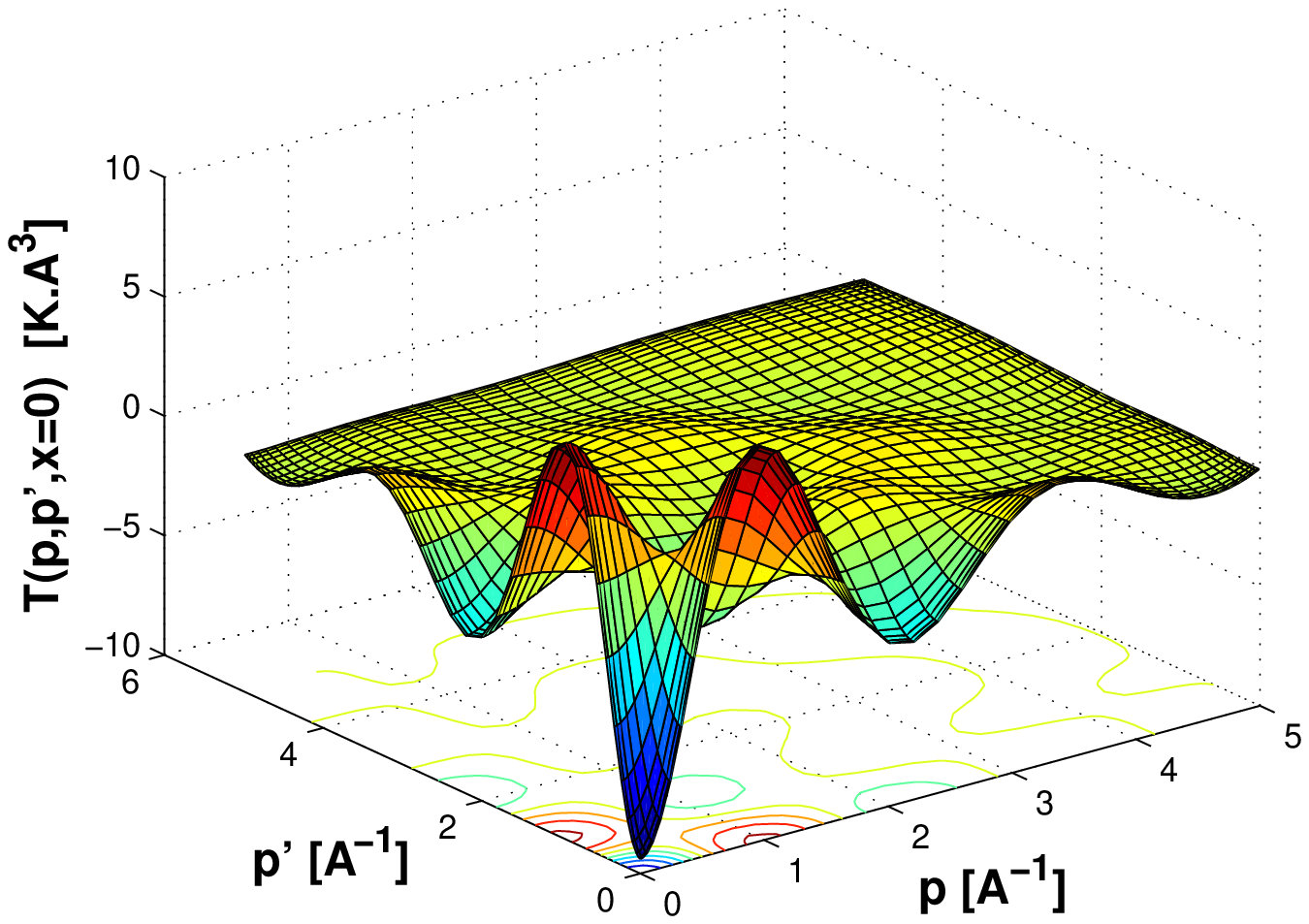} &
\includegraphics*[width=5.2cm]{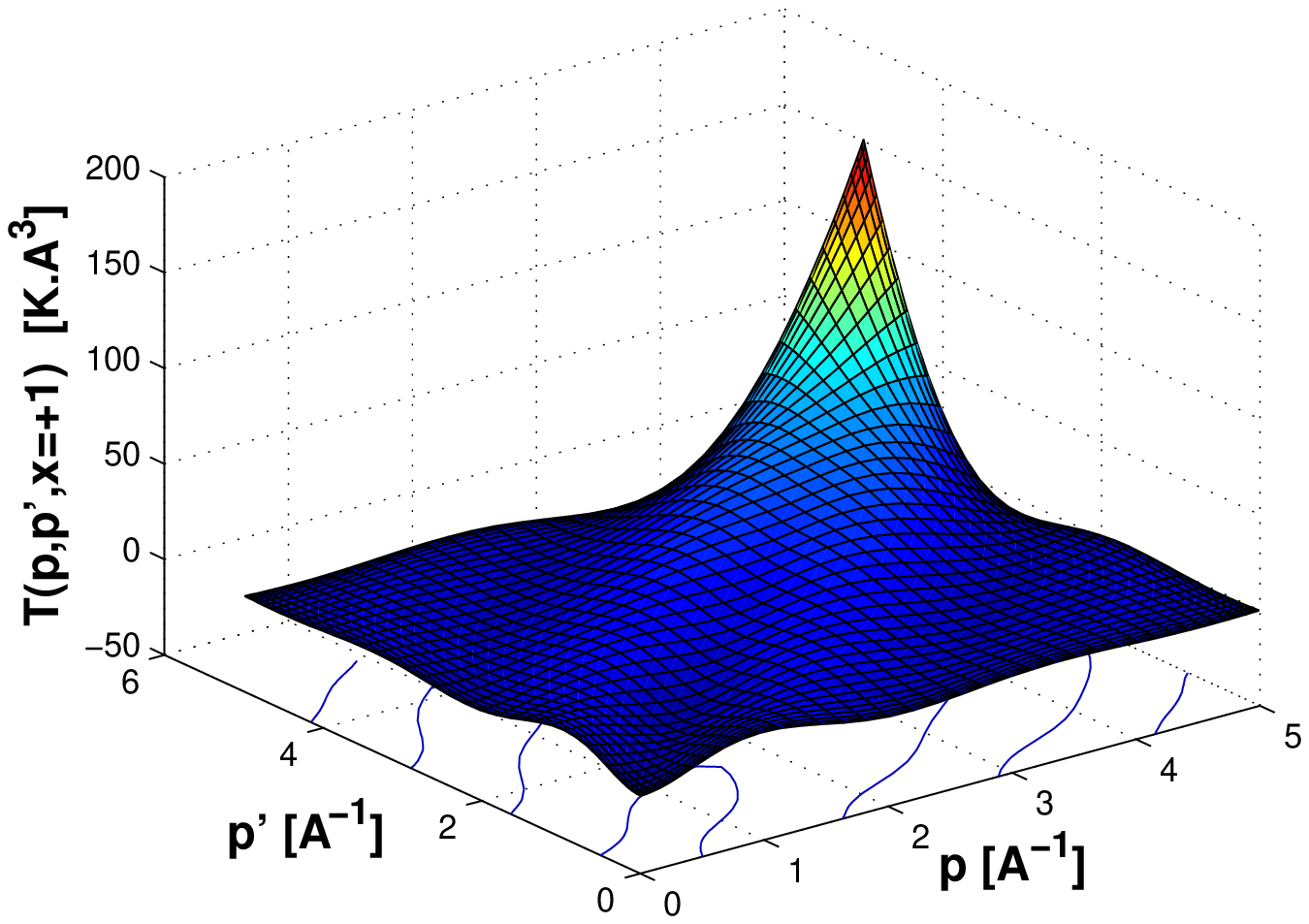} \\
HFDHE2 & HFDHE2 & HFDHE2 \\
\includegraphics*[width=5.2cm]{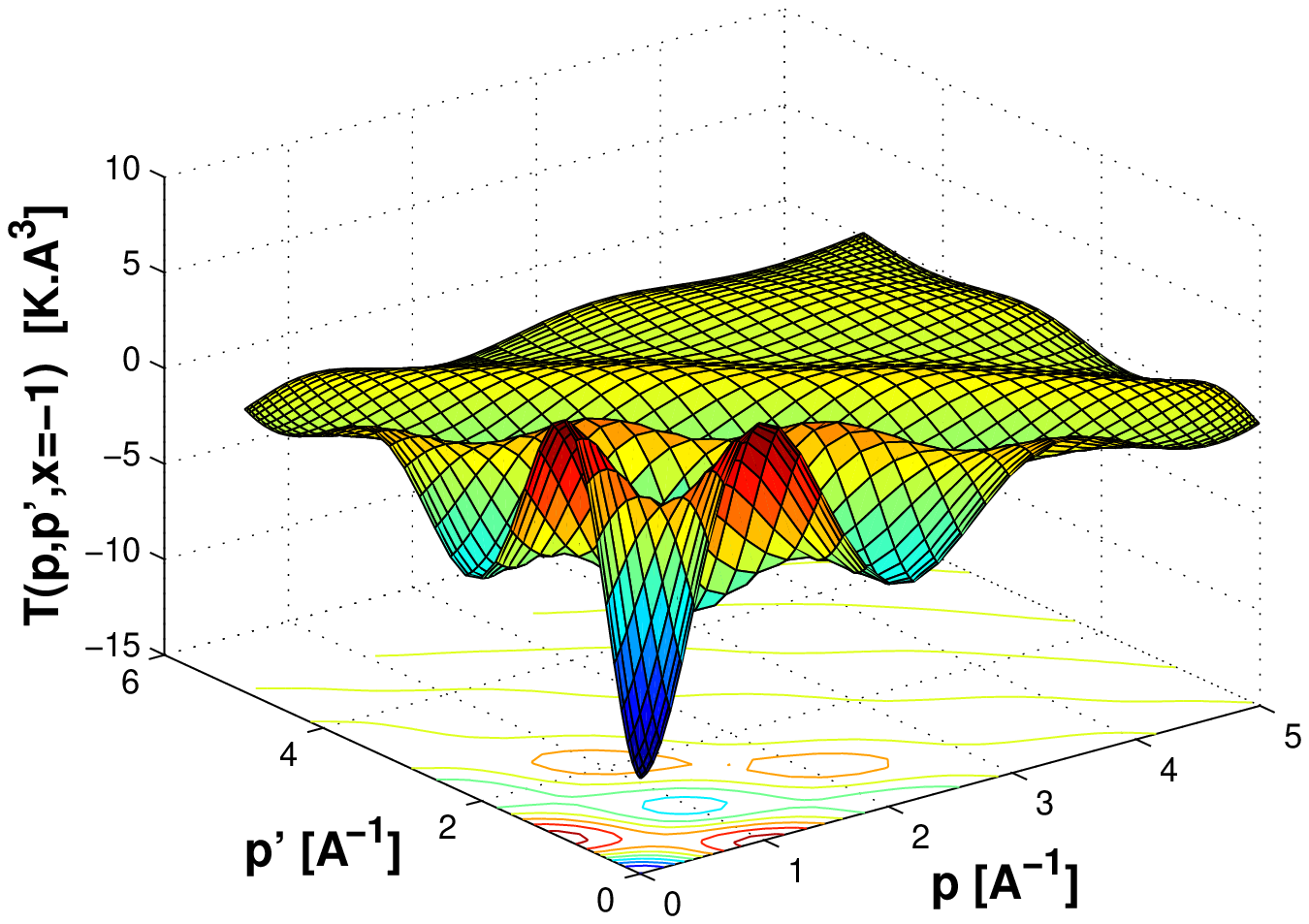} &
\includegraphics*[width=5.2cm]{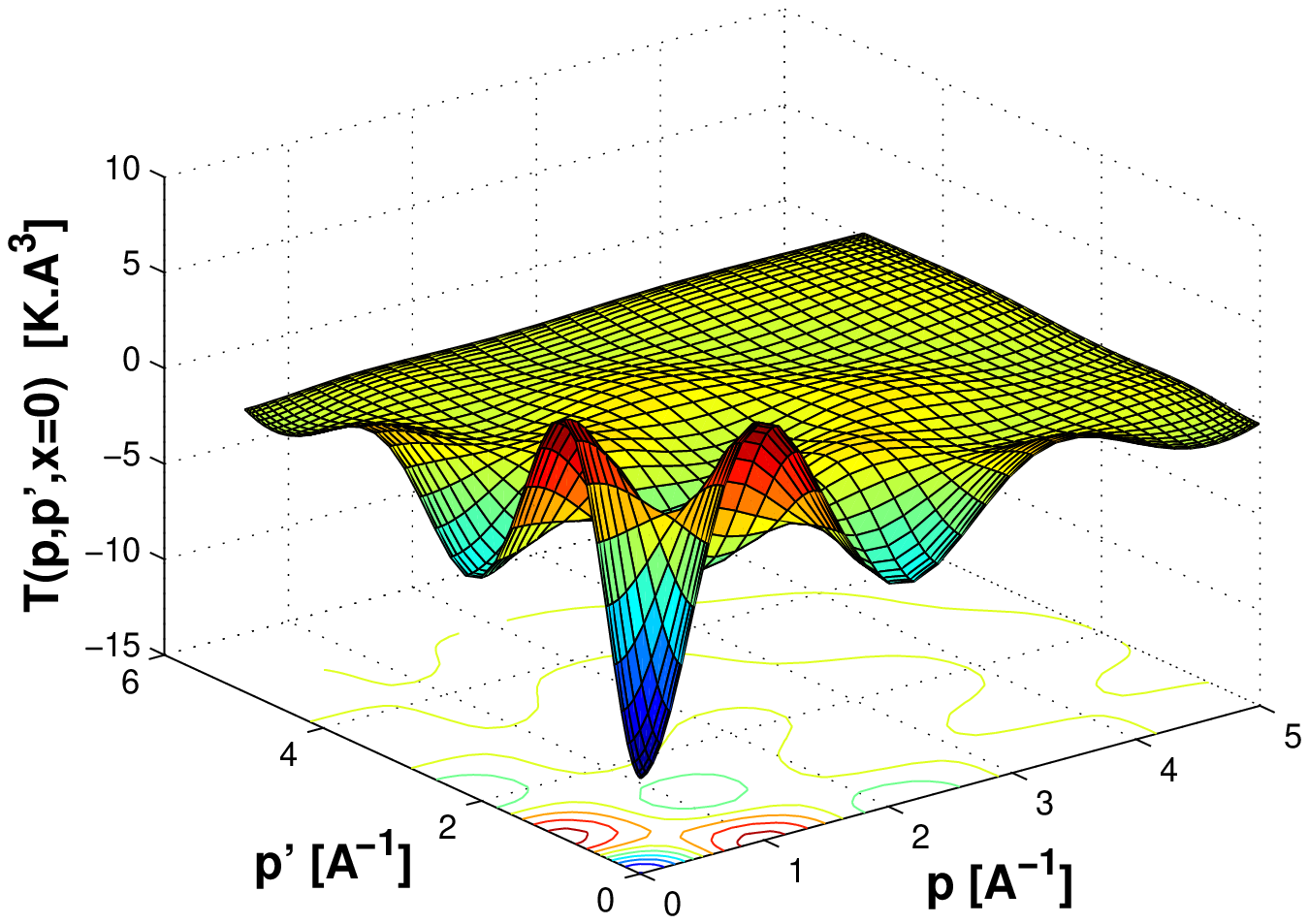} &
\includegraphics*[width=5.2cm]{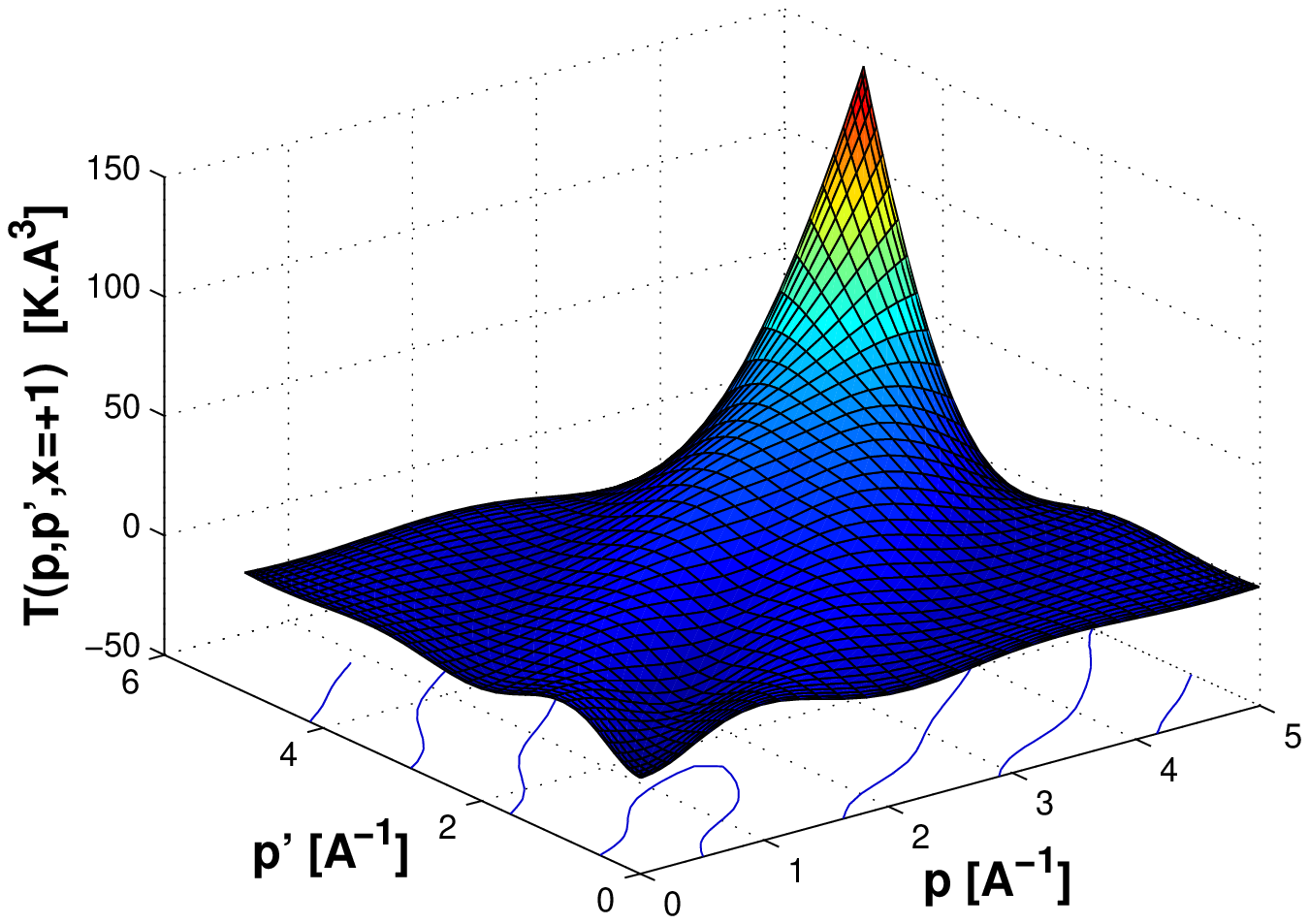} \\
HFD-B & HFD-B & HFD-B \\
\includegraphics*[width=5.2cm]{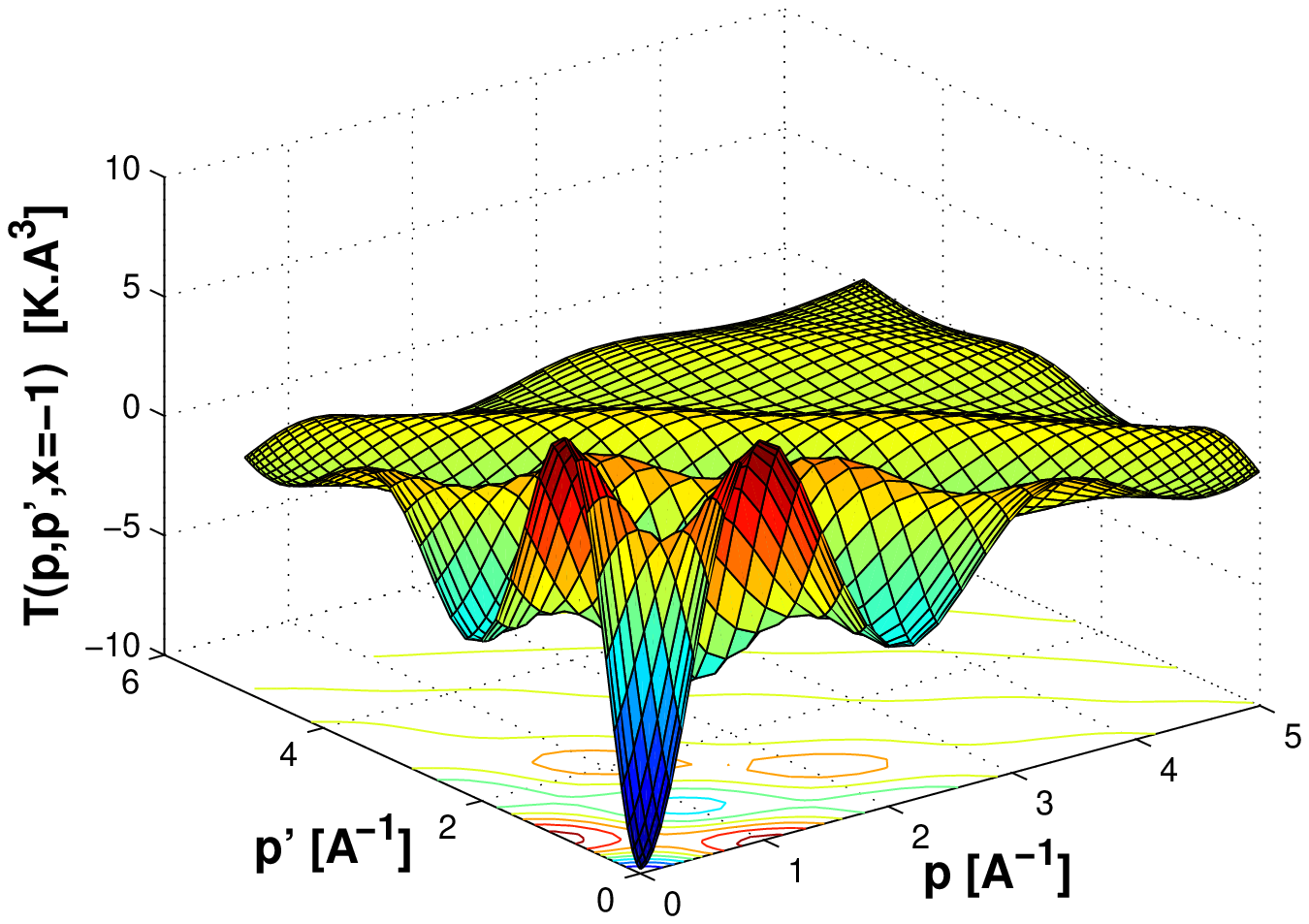} &
\includegraphics*[width=5.2cm]{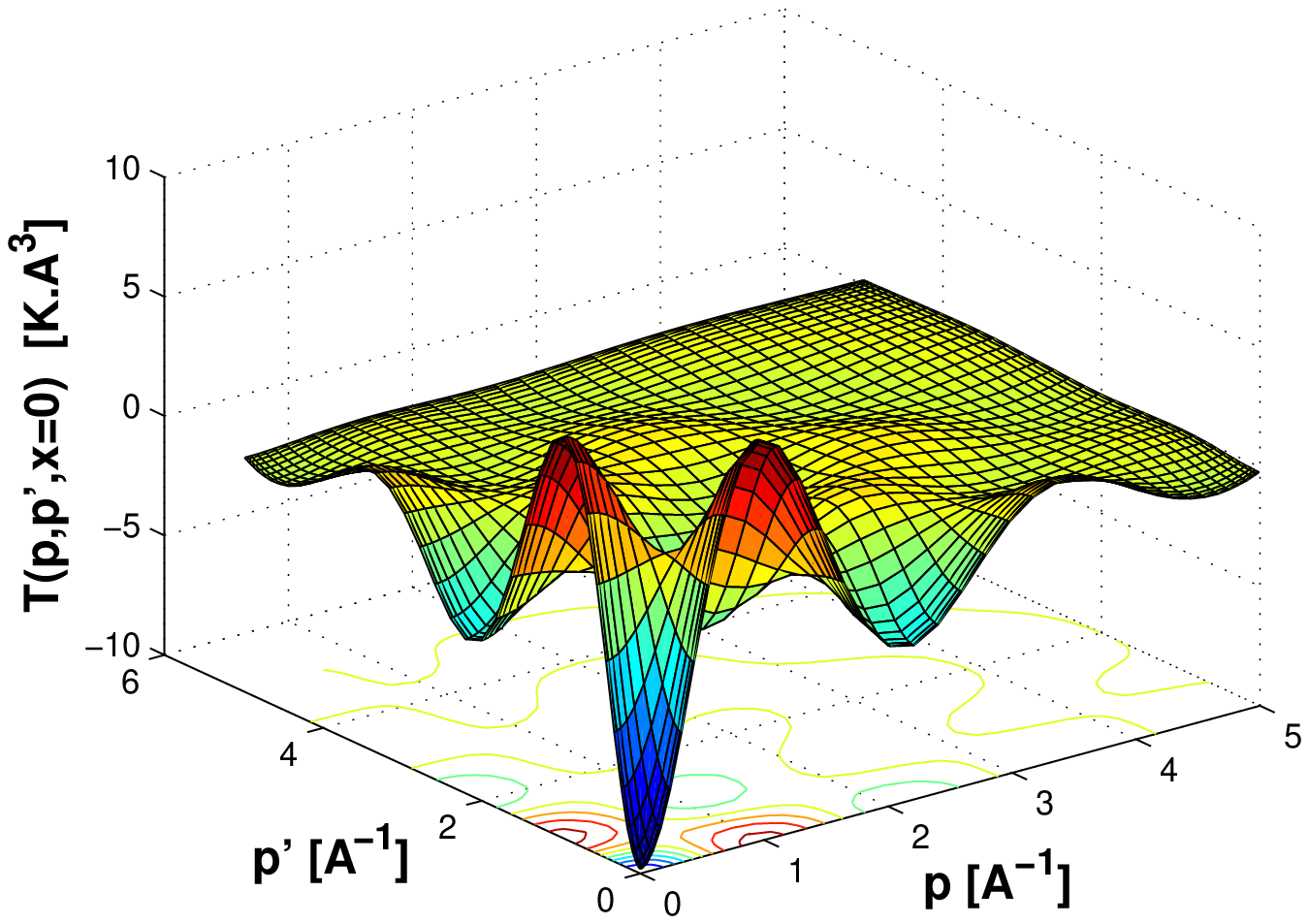} &
\includegraphics*[width=5.2cm]{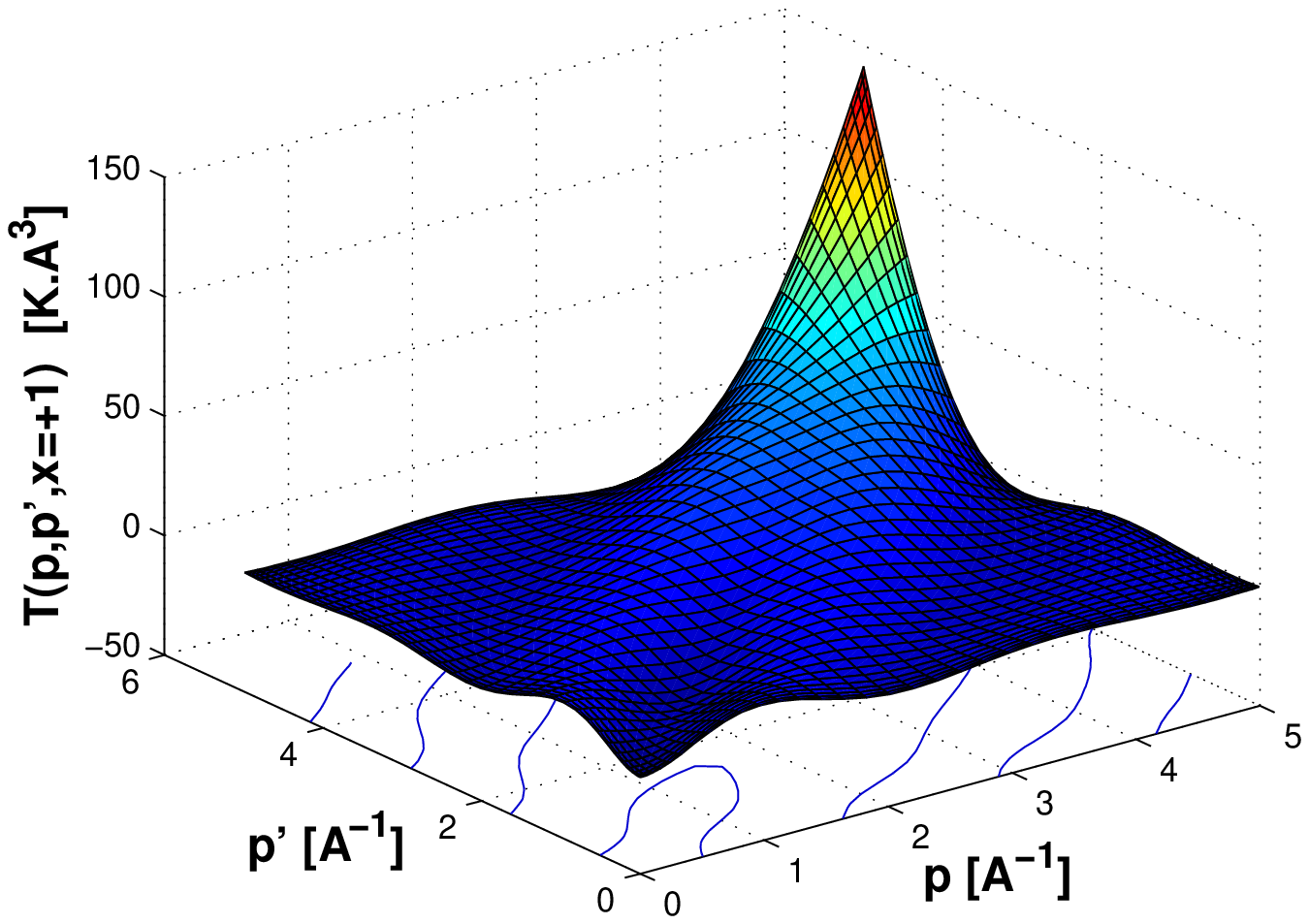} \\
LM2M2 & LM2M2 & LM2M2 \\
\includegraphics*[width=5.2cm]{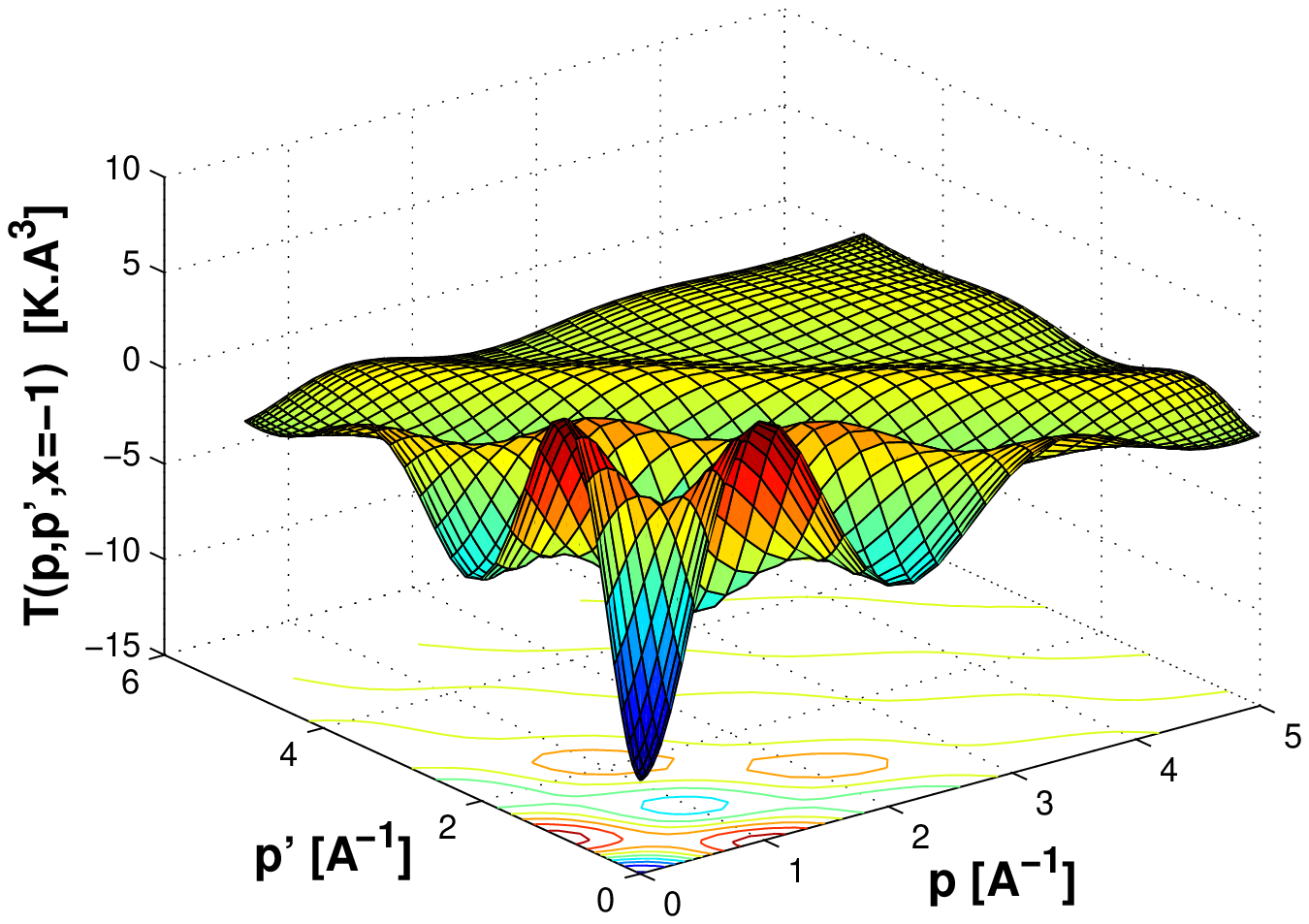} &
\includegraphics*[width=5.2cm]{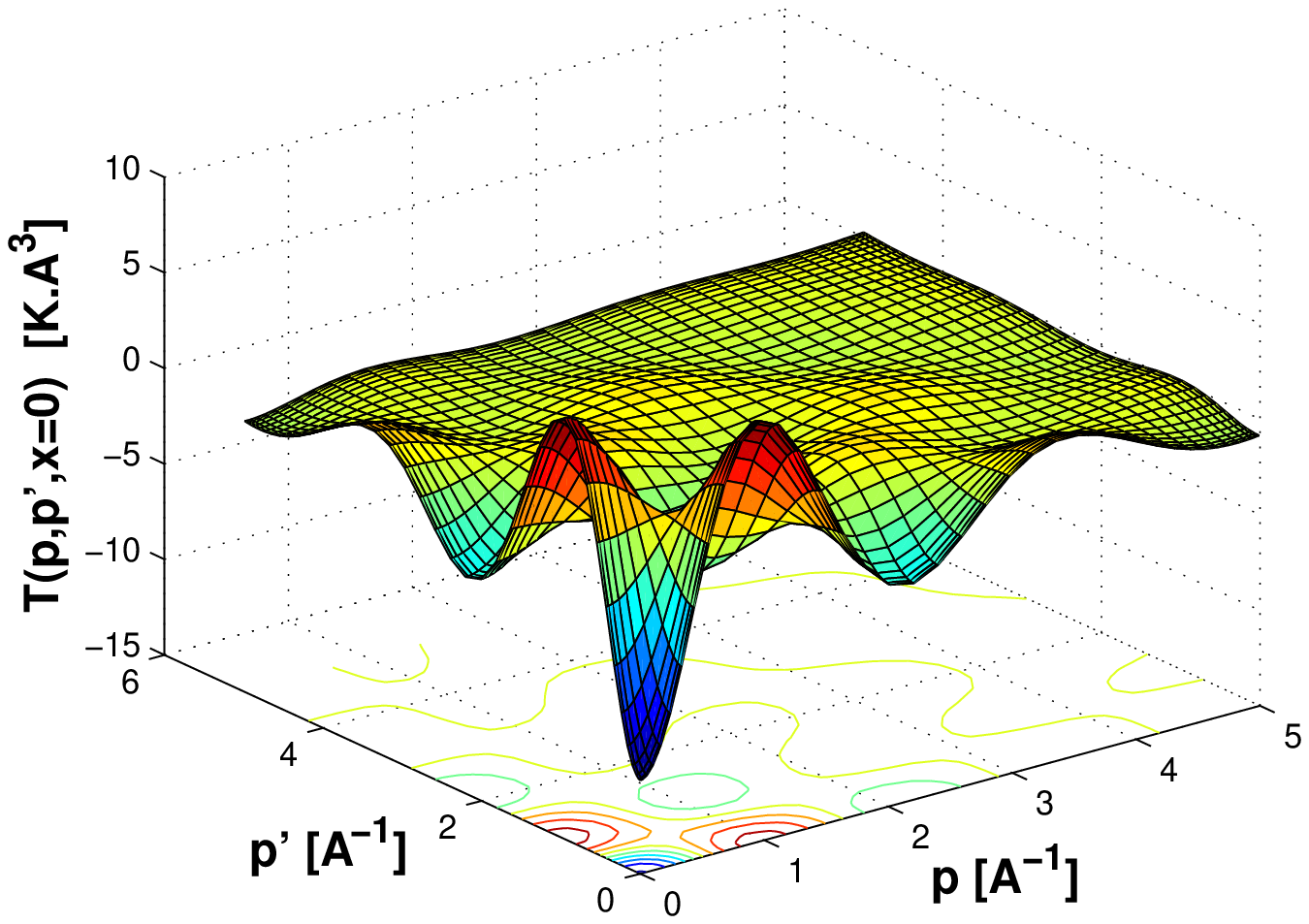} &
\includegraphics*[width=5.2cm]{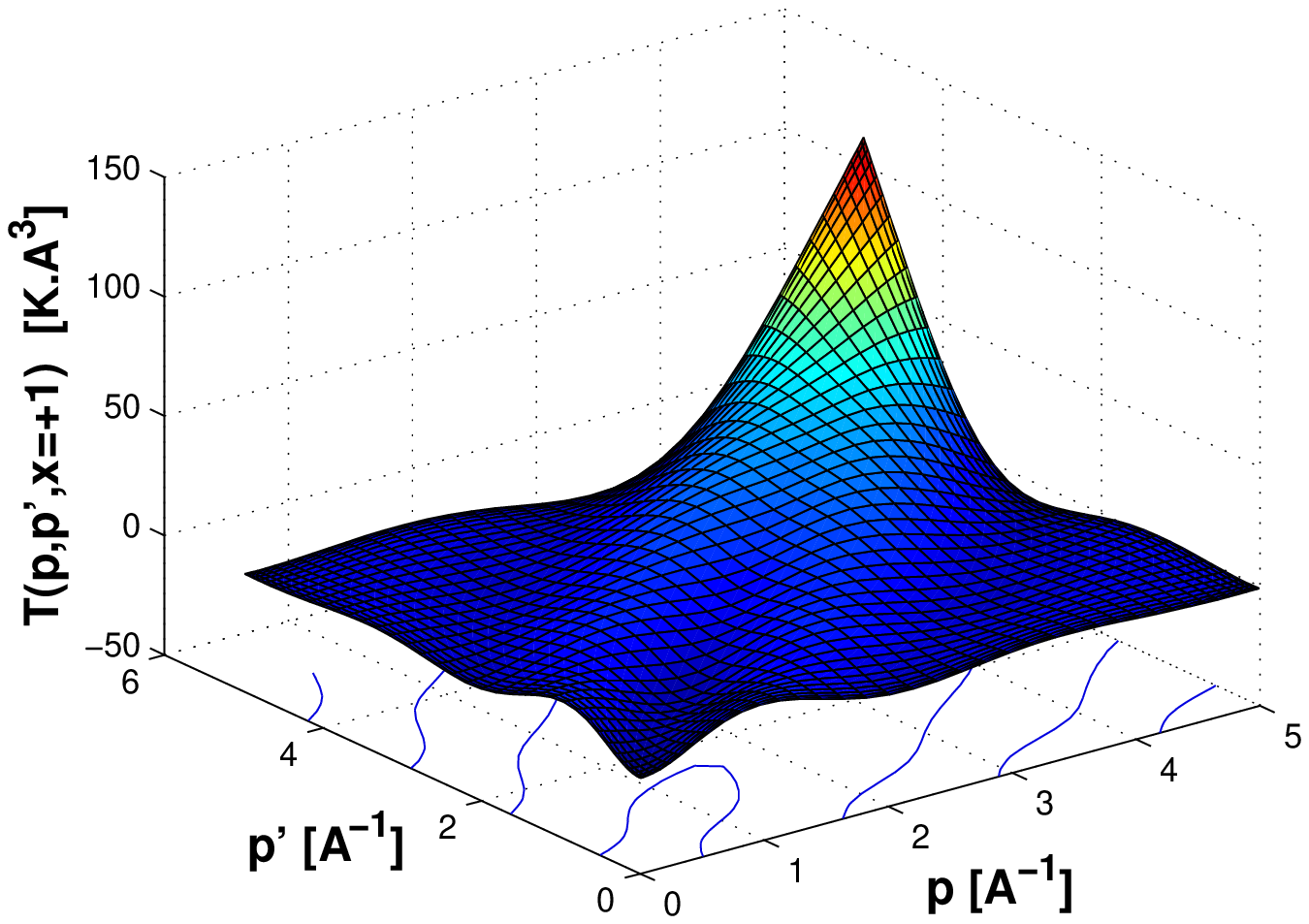} \\
TTY & TTY & TTY
\end{tabular}
\caption{Momentum dependence of the fully offshell transition amplitude $T(p,p',x_{pp'};E)$ at $E=-100 \, mK$ in fixed angles $x_{pp'}=0,\pm1$.} \label{fig.fully-of-shell-t-matrix}
\end{figure*}

\begin{figure*}[hbt]
\centering
\begin{tabular}{cc}
\includegraphics*[width=6cm]{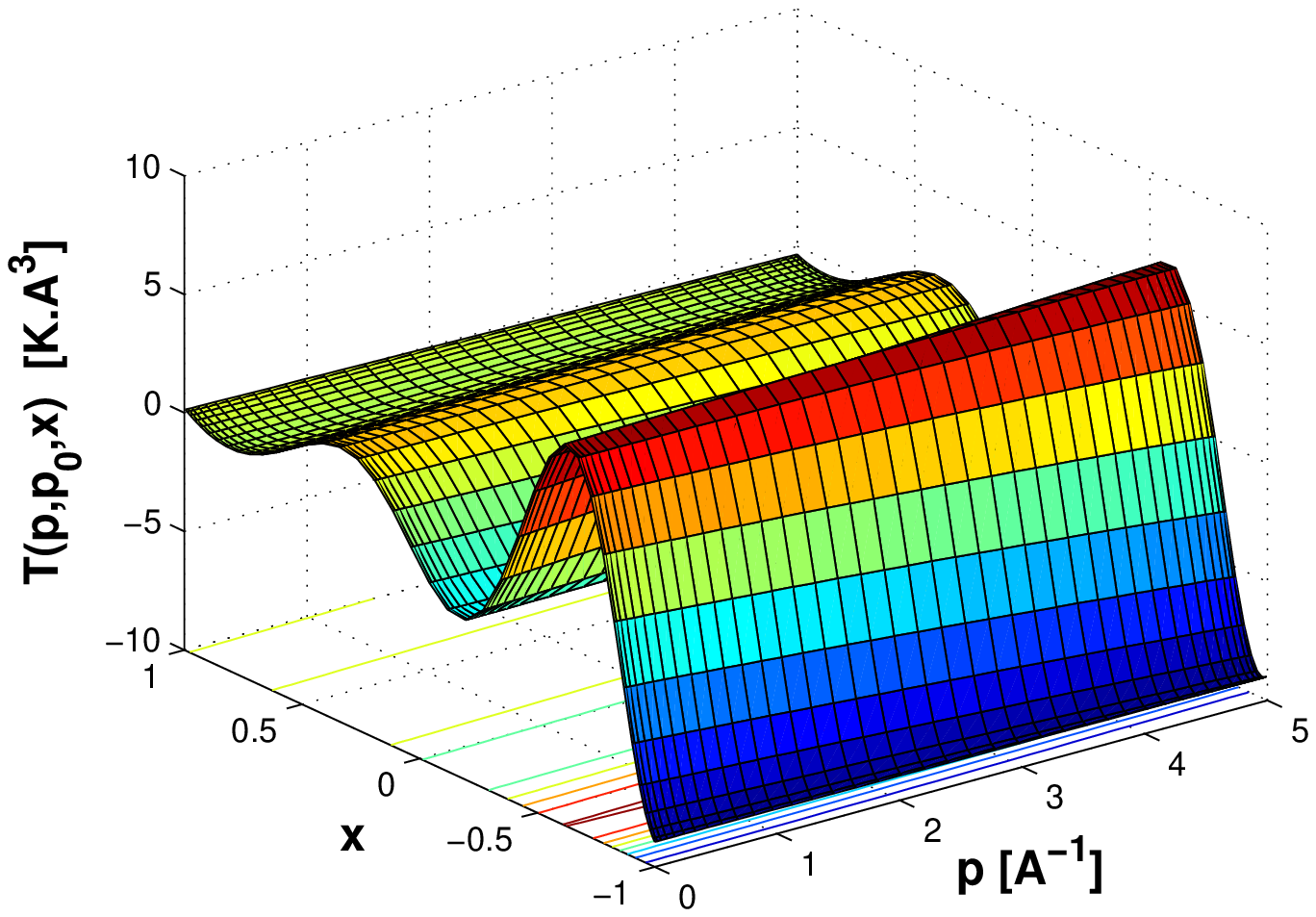} &
\includegraphics*[width=6cm]{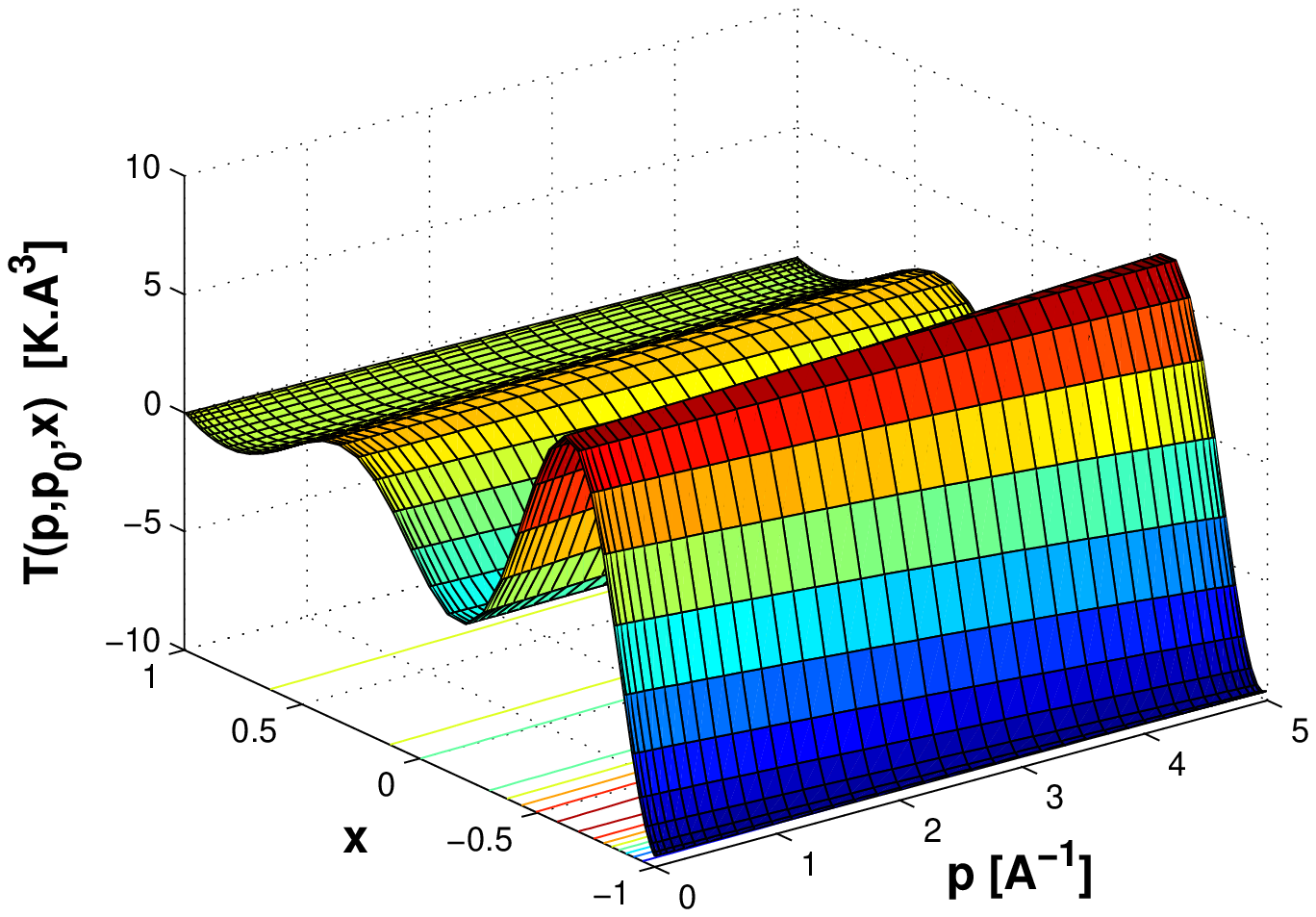} \\
HFDHE2 & HFD-B \\
\includegraphics*[width=6cm]{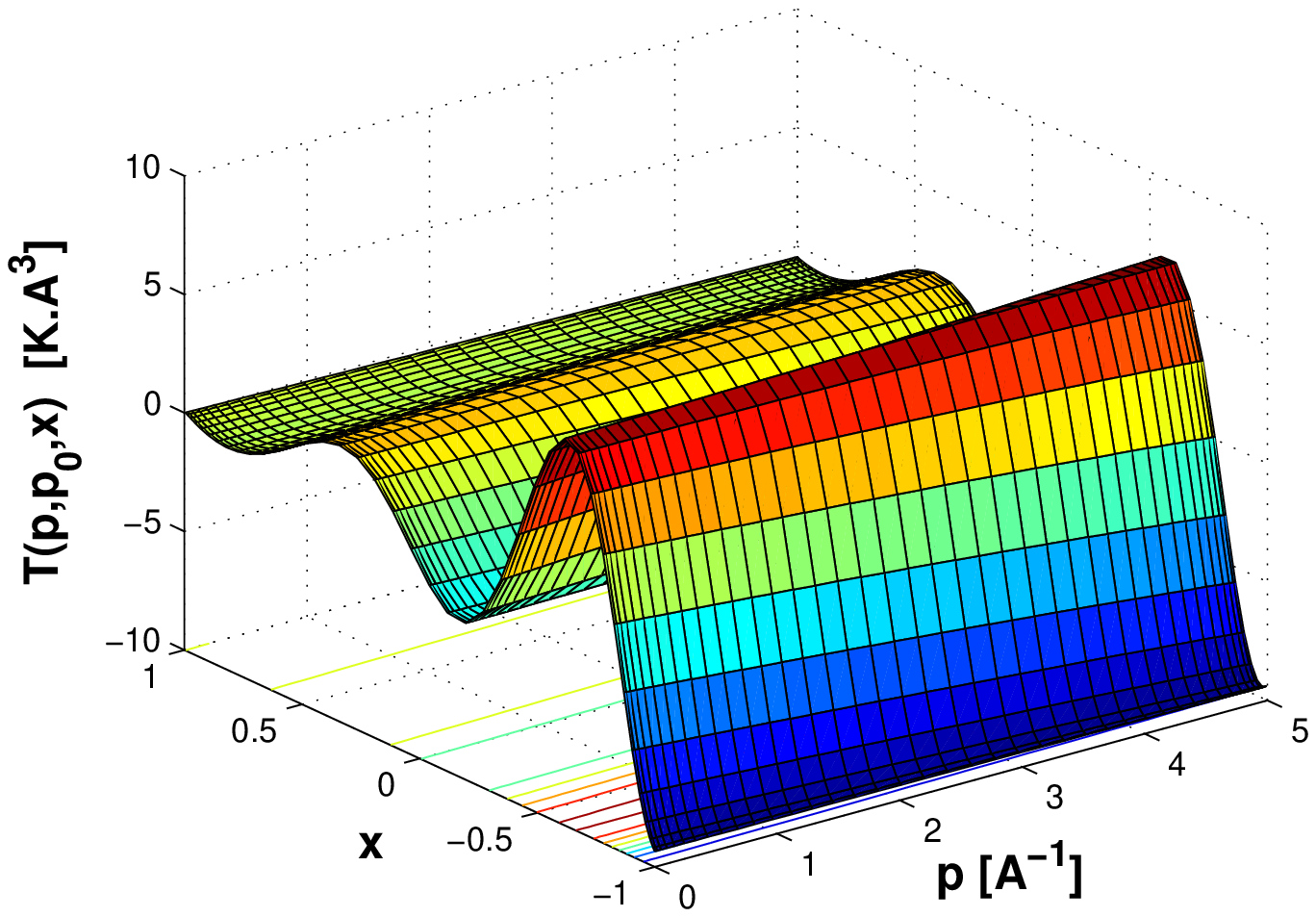} &
\includegraphics*[width=6cm]{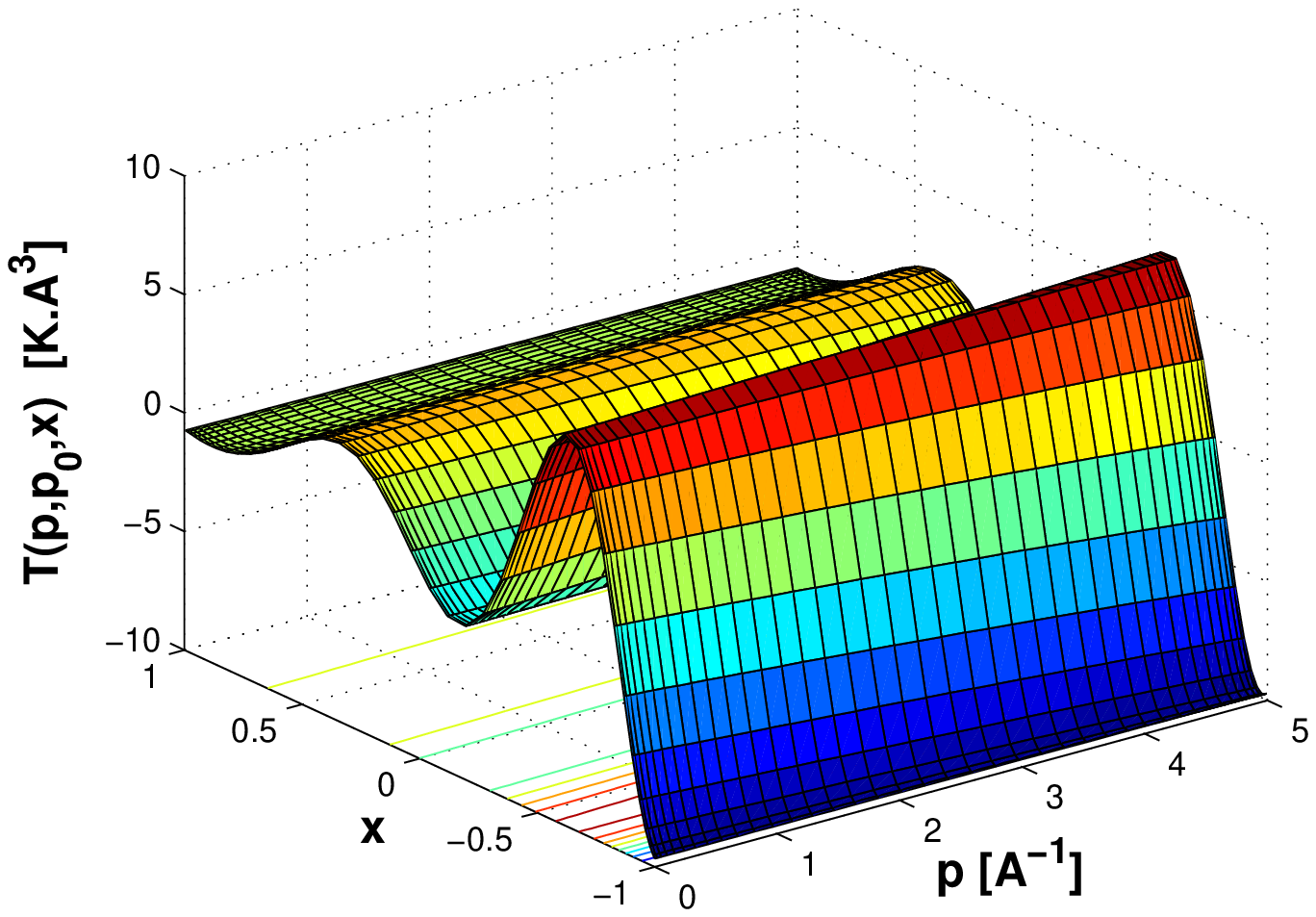} \\
LM2M2 & TTY
\end{tabular}
\caption{Momentum and angle dependences of $T(p,p_0,x_{pp'})$ with $p_0=\sqrt{m|E|}$ at $E=-100 \, mK$. } \label{fig.half-of-shell-t-matrix}
\end{figure*}

For a specific value of the off-shell momentum $p$ and energy $E$ and after discretization of continuous momentum and angle variables, this two-dimensional integral equation can be turned into a system of linear equations as $AT=B$, where $A$ and $B$ are composed of kernel of integral equation and potential matrix elements respectively. For our numerical calculations we use the Lapack Fortran library \cite{Lapack} to solve the obtained
system of linear equations. Certainly for few-atomic scattering state studies one needs to calculate the transition amplitude at positive energies which leads to a singularity in free propagator. This Cauchy singularity can be splitted easily into a principal-value integral and a $\delta$-function imaginary part. Since we are going to use the recently developed formalism for three- and four-body bound states \cite{Elster-FBS27}-\cite{Hadizadeh-EPJA36} for $^4$He trimer and tetramer calculations we study the behavior of transition amplitude at negative energies.
To this aim we have solved the two-dimensional integral equation (\ref{eq.T-final}) by using 40 and 41 mesh grids for Jacobi momenta and spherical angles variables respectively. We would like to mention that by considering the symmetry property of the angle argument $x_{p'\,p''}$, the polar angle integration in Eq. (\ref{eq.V-phi-integration}) can be done on interval $[0,\frac{\pi}{2}]$ by using 10 mesh grids. Our numerical results for fully offshell transition amplitudes $T(p,p',x_{pp'};E)$ are shown in Figure (\ref{fig.fully-of-shell-t-matrix}) at energy $E=-100\, mK$, which is close to $^4$He trimer binding energy, in fixed angles $x_{pp'}=0,\pm1$. In Figure (\ref{fig.half-of-shell-t-matrix}) we have shown the momentum and angular dependence of half offshell transition amplitude $T(p,p_0=\sqrt{m|E|},x;E)$ at energy $E=-100\, mK$.

\begin{figure*}[hbt]
\centering
\begin{tabular}{cc}
\includegraphics*[width=6cm]{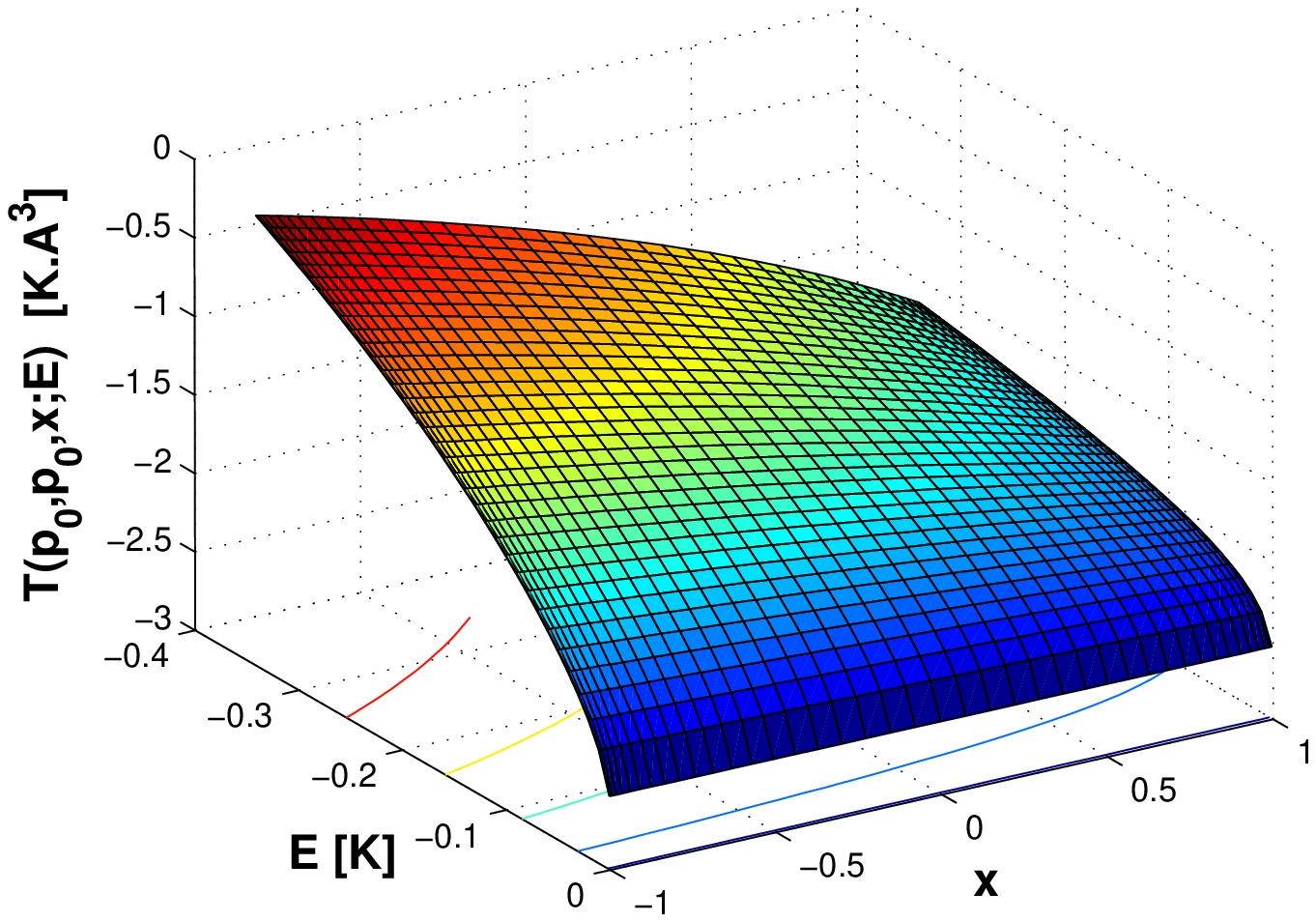} &
\includegraphics*[width=6cm]{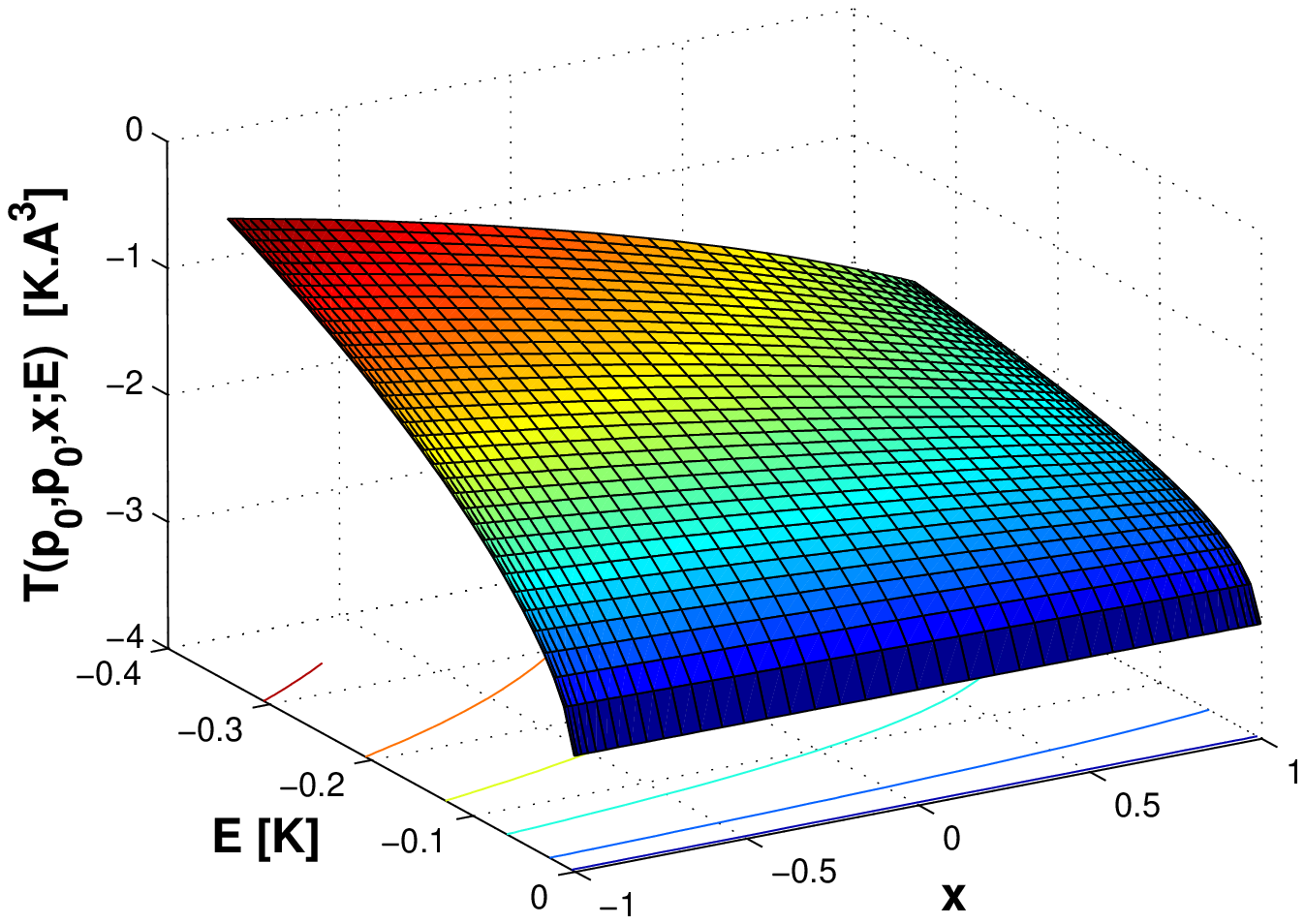} \\
HFDHE2 & HFD-B \\
\includegraphics*[width=6cm]{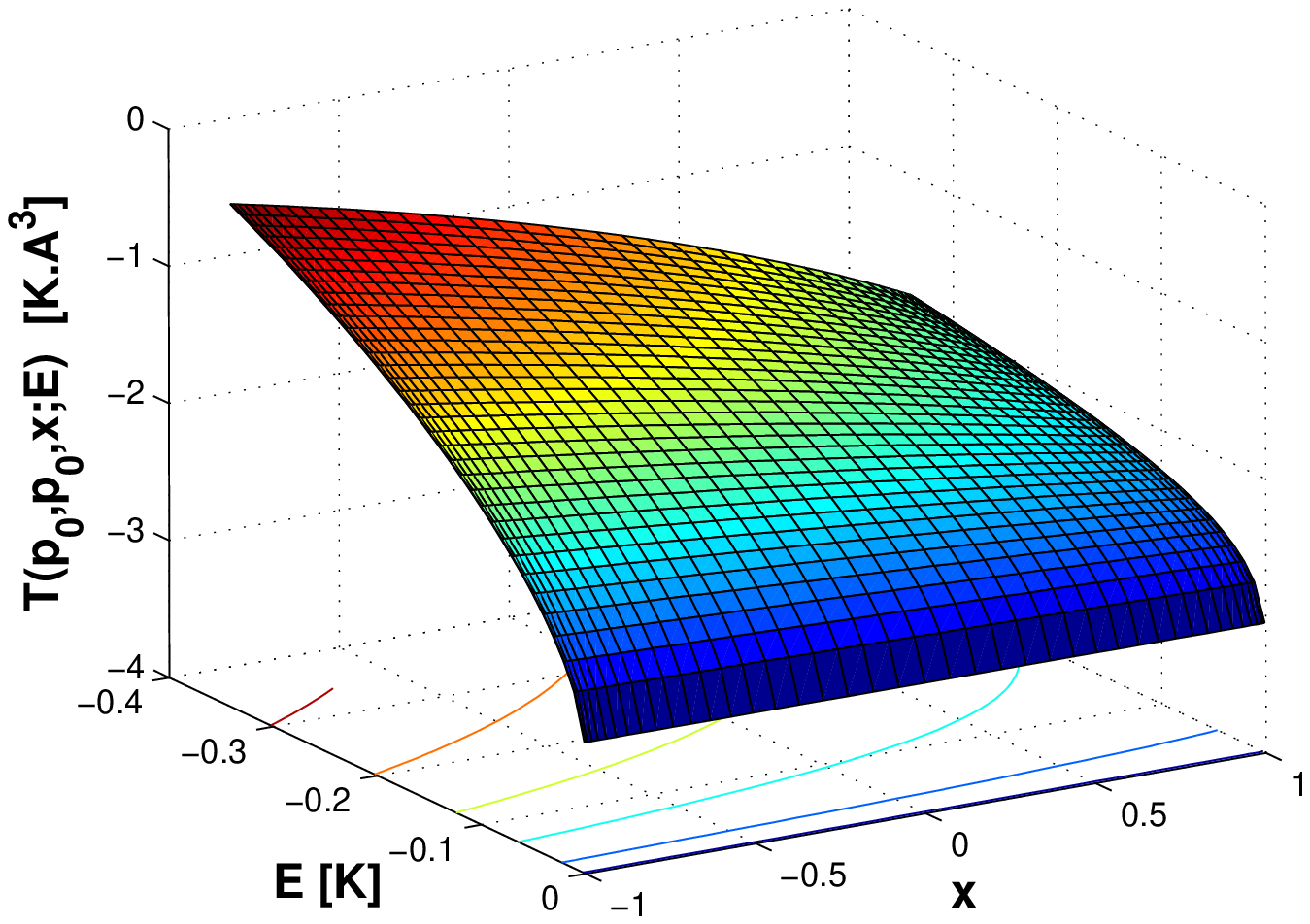} &
\includegraphics*[width=6cm]{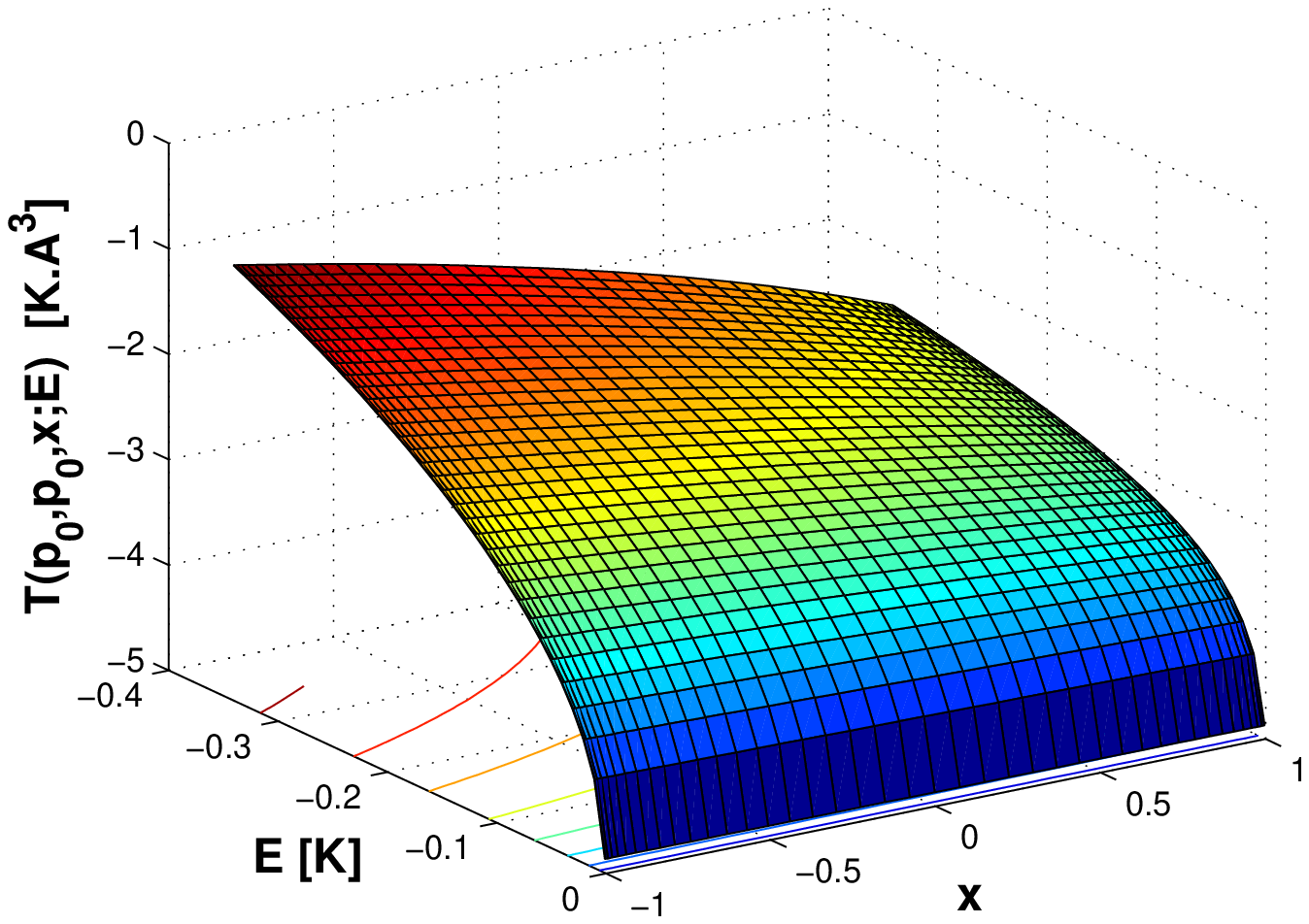}  \\
LM2M2 & TTY \\
\end{tabular}
\caption{Angular dependence of $T(p_0,p_0,x_{pp'};E)$ with $p_0=\sqrt{m|E|}$ as function of the energy from $E=-1 \, mK$ to $E=-400 \, mK$.} \label{fig.T-matrix-pole-Energy-range}
\end{figure*}

\begin{figure*}[hbt]
\centering
\begin{tabular}{cc}
\includegraphics*[width=6cm]{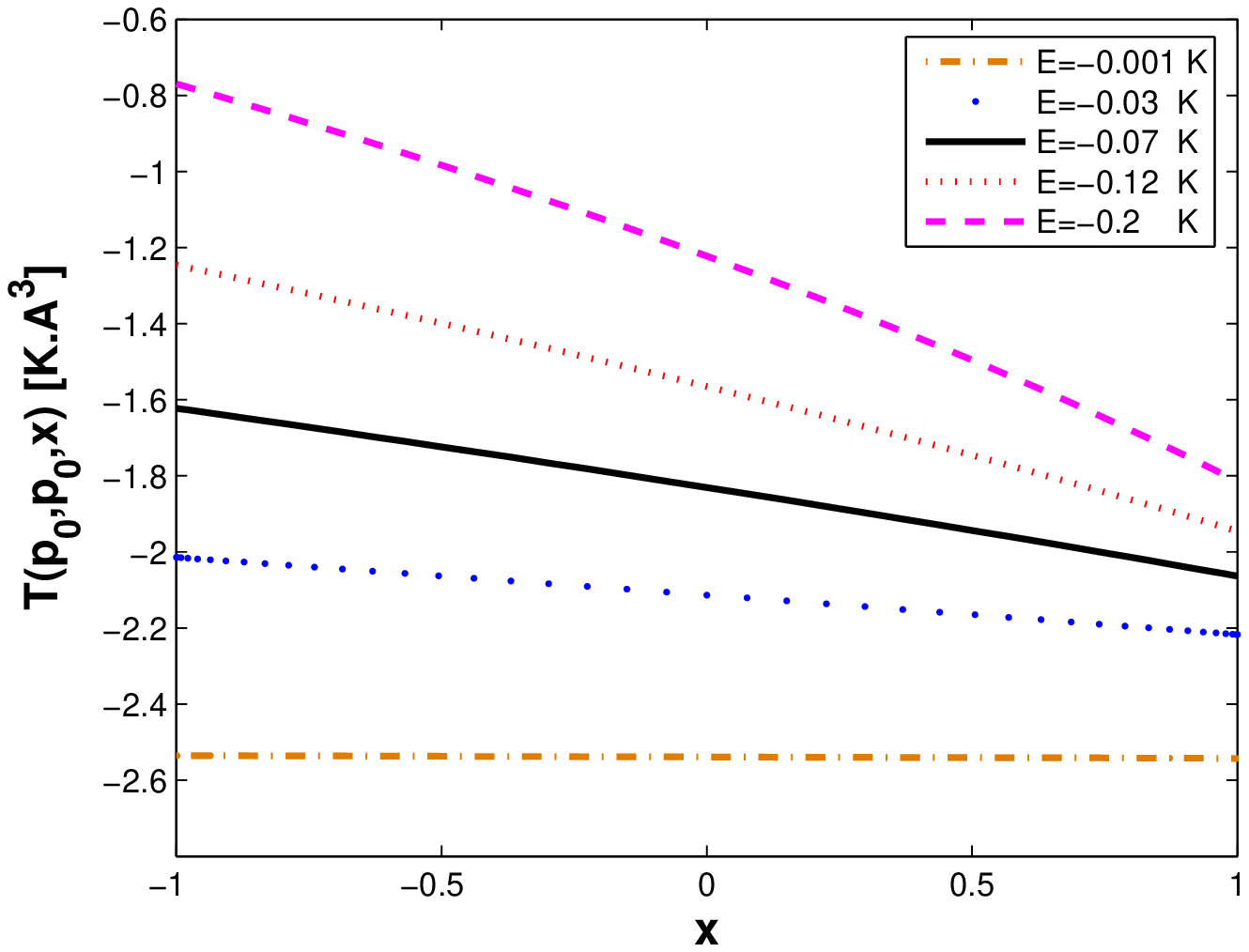} &
\includegraphics*[width=6cm]{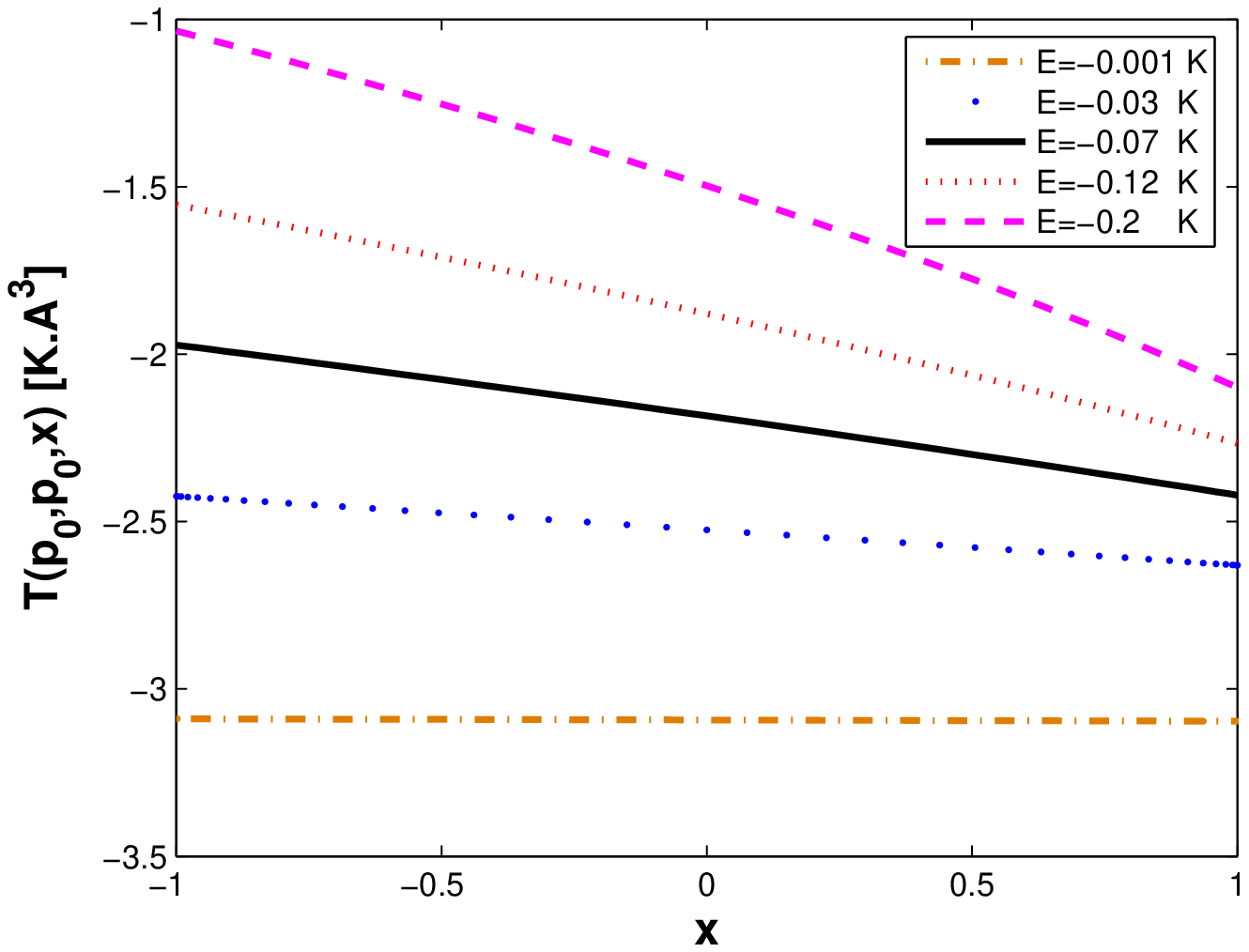} \\
HFDHE2 & HFD-B \\
\includegraphics*[width=6cm]{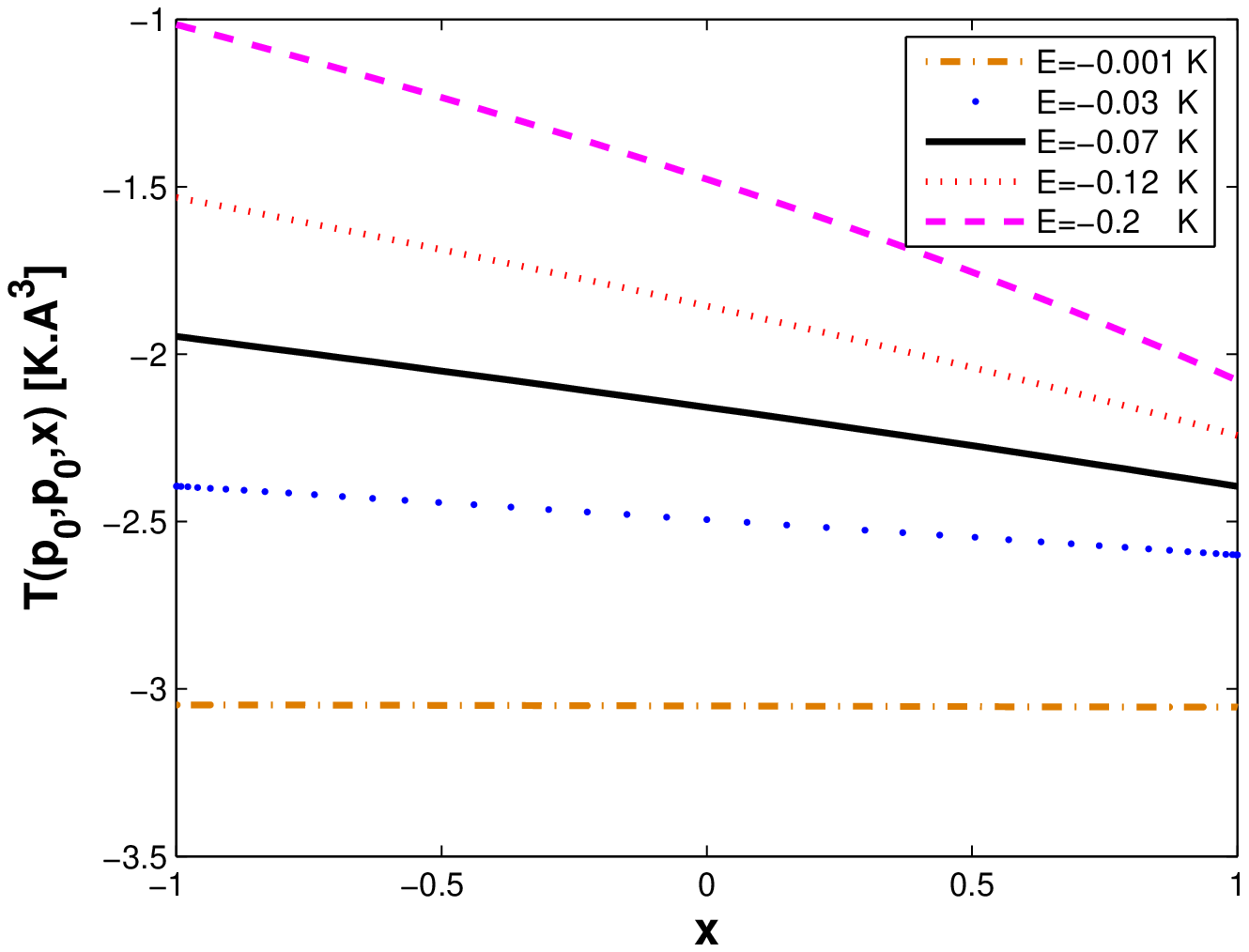} &
\includegraphics*[width=6cm]{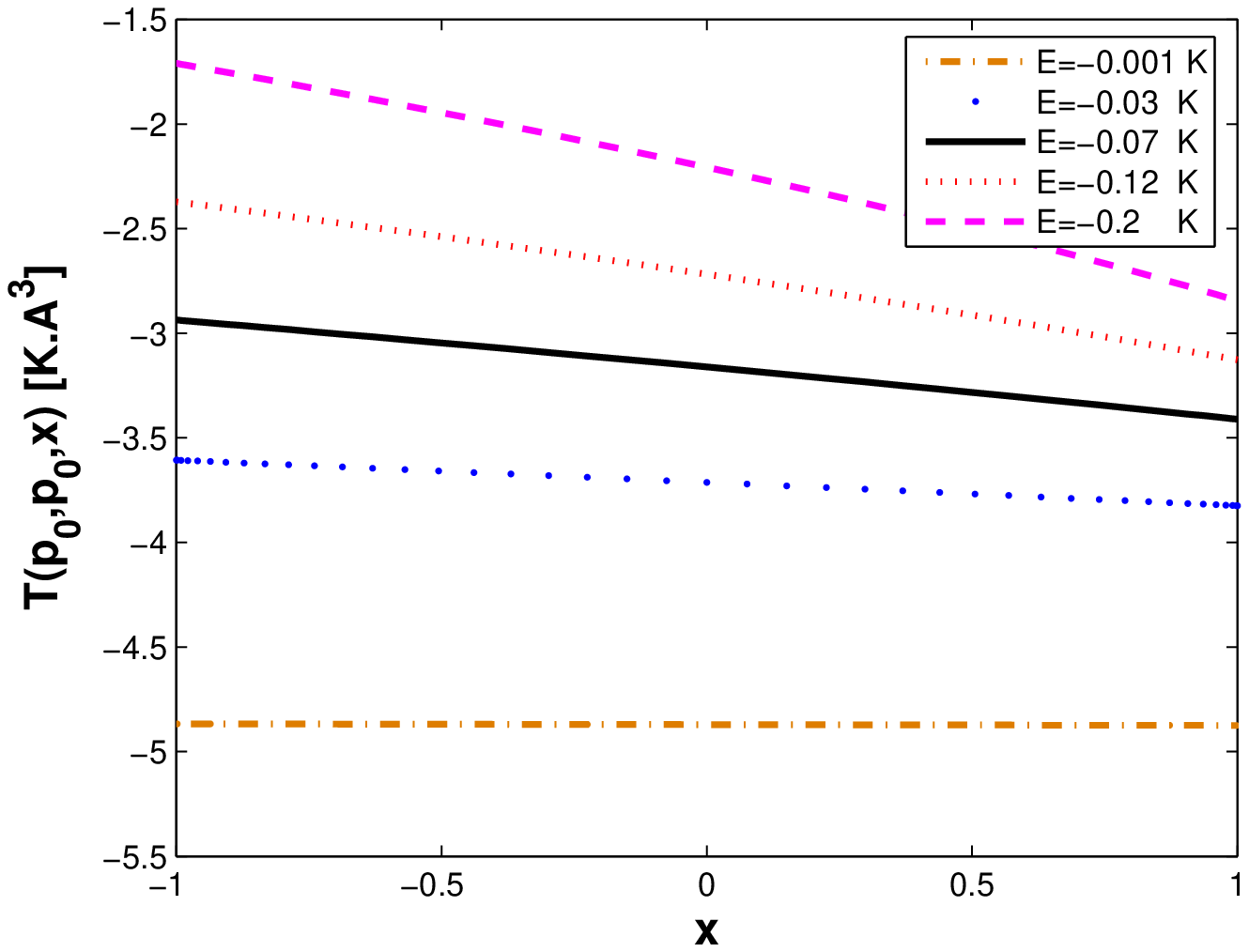}  \\
LM2M2 & TTY \\
\end{tabular}
\caption{Angular dependence of $T(p_0,p_0,x_{pp'};E)$ with $p_0=\sqrt{m|E|}$ at energies around $^4$He dimer $s$-wave pole.} \label{fig.T-matrix-pole}
\end{figure*}

As shown in Ref. \cite{Elster-FBS24} the bound states of two-body system lead to poles in the transition amplitude and the angular dependence of transition amplitude exhibits a very characteristic behavior in the vicinity of the bound state poles, which is given by the Legendre function corresponding to angular quantum number of the bound state
\begin{eqnarray}
T({\bf p'},{\bf p};E) & \stackrel{E \rightarrow E_b}{\longrightarrow}& \frac {2l+1}{4\pi} P_l(x_{pp'})
\frac {g_l(p') \, g_l(p)} {E-E_b},
 \label{eq.3.5}
\end{eqnarray}
where
\begin{eqnarray}
g_l(p) = \int_0^{\infty} dp' p'^2 \, v_l(p,p') \, \psi_{l}(p'),
\label{eq.3.6}
\end{eqnarray}
where $v_l(p,p')$ and $\psi_{l}(p')$ are partial wave components of potential and tw-body wave function.
To investigate this characteristic behavior in atomic case we have shown in Figure (\ref{fig.T-matrix-pole-Energy-range}) the angular dependence of transition amplitude $T(p_0,p_0,x;E)$ at the energy range $-400 \leq E \leq -1 \, mK$. Clearly we can see by decreasing the magnitude of energy the angular behavior of transition amplitude is corresponding to the zeroth Legendre polynomial, i.e. $P_0(x_{pp'})$. For a better representation of this angular behavior we have shown in Figure (\ref{fig.T-matrix-pole}) the transition amplitude just for few energies close to dimer $s-$wave pole and also the magnitudes of transition amplitude at energy $E=-1 \, mK$ are listed in Table (\ref{table.pole}).

\begin{table}[hbt]
\caption {The matrix elements of $T(p_0,p_0,x_{pp'};E)$ in unit of $K.\AA^3$ with $p_0=\sqrt{m|E|}$ at energy $E=-1 \, mK$, which is close to $s-$wave $^4$He dimer pole.}
\label{table.pole}
\begin{tabular}{ccccccccccccccccccccccccccccccccccccccccccccccccc}
\hline\noalign{\smallskip}
 $x_{pp'}=\hat{p}.\hat{p'}$ &   HFDHE2   & HFD-B    & LM2M2     & TTY \\ \hline
-1.00000	&	-2.5353	&	-3.0891	&	-3.0473	&	-4.8667	\\
-0.99832	&	-2.5353	&	-3.0892	&	-3.0473	&	-4.8667	\\
-0.99117	&	-2.5353	&	-3.0892	&	-3.0473	&	-4.8668	\\
-0.97834	&	-2.5353	&	-3.0892	&	-3.0473	&	-4.8668	\\
-0.95991	&	-2.5354	&	-3.0893	&	-3.0474	&	-4.8669	\\
-0.93598	&	-2.5355	&	-3.0894	&	-3.0475	&	-4.8670	\\
-0.90669	&	-2.5356	&	-3.0895	&	-3.0476	&	-4.8671	\\
-0.87220    &	-2.5357	&	-3.0896	&	-3.0477	&	-4.8672	\\
-0.83272	&	-2.5359	&	-3.0898	&	-3.0479	&	-4.8674	\\
-0.78847	&	-2.5360	&	-3.0899	&	-3.0480	&	-4.8676	\\
-0.73970    &	-2.5362	&	-3.0901	&	-3.0482	&	-4.8677	\\
-0.68670    &	-2.5364	&	-3.0903	&	-3.0484	&	-4.8679	\\
-0.62977	&	-2.5366	&	-3.0905	&	-3.0486	&	-4.8682	\\
-0.56922	&	-2.5368	&	-3.0907	&	-3.0488	&	-4.8684	\\
-0.50542	&	-2.5370	&	-3.0909	&	-3.0491	&	-4.8686	\\
-0.43872	&	-2.5373	&	-3.0912	&	-3.0493	&	-4.8689	\\
-0.36950    &	-2.5375	&	-3.0914	&	-3.0495	&	-4.8692	\\
-0.29818	&	-2.5378	&	-3.0917	&	-3.0498	&	-4.8694	\\
-0.22514	&	-2.5380	&	-3.0920	&	-3.0501	&	-4.8697	\\
-0.15081	&	-2.5383	&	-3.0922	&	-3.0503	&	-4.8700	\\
-0.07562	&	-2.5386	&	-3.0925	&	-3.0506	&	-4.8703	\\
2.47E-32	&	-2.5388	&	-3.0928	&	-3.0509	&	-4.8706	\\
+0.07562	&	-2.5391	&	-3.0931	&	-3.0512	&	-4.8709	\\
+0.15081    &	-2.5394	&	-3.0933	&	-3.0514	&	-4.8711	\\
+0.22514    &	-2.5397	&	-3.0936	&	-3.0517	&	-4.8714	\\
+0.29818    &	-2.5399	&	-3.0939	&	-3.0520	&	-4.8717	\\
+0.36950    &	-2.5402	&	-3.0941	&	-3.0522	&	-4.8720	\\
+0.43872    &	-2.5404	&	-3.0944	&	-3.0525	&	-4.8723	\\
+0.50542    &	-2.5407	&	-3.0946	&	-3.0527	&	-4.8725	\\
+0.56922    &	-2.5409	&	-3.0949	&	-3.0530	&	-4.8728	\\
+0.62977    &	-2.5411	&	-3.0951	&	-3.0532	&	-4.8730	\\
+0.68670    &	-2.5413	&	-3.0953	&	-3.0534	&	-4.8732	\\
+0.73970    &	-2.5415	&	-3.0955	&	-3.0536	&	-4.8734	\\
+0.78847    &	-2.5417	&	-3.0957	&	-3.0538	&	-4.8736	\\
+0.83272    &	-2.5418	&	-3.0958	&	-3.0539	&	-4.8738	\\
+0.87220    &	-2.5420	&	-3.0960	&	-3.0541	&	-4.8739	\\
+0.90669    &	-2.5421	&	-3.0961	&	-3.0542	&	-4.8740	\\
+0.93598    &	-2.5422	&	-3.0962	&	-3.0543	&	-4.8742	\\
+0.95991    &	-2.5423	&	-3.0963	&	-3.0544	&	-4.8743	\\
+0.97834    &	-2.5424	&	-3.0963	&	-3.0544	&	-4.8743	\\
+0.99117    &	-2.5424	&	-3.0964	&	-3.0545	&	-4.8744	\\
+0.99832    &	-2.5424	&	-3.0964	&	-3.0545	&	-4.8744	\\
+1.00000    &	-2.5424	&	-3.0964	&	-3.0545	&	-4.8744	\\
\noalign{\smallskip}\hline
\end{tabular}
\end{table}

\section{Outlook}\label{sec.outlook}

The first step toward the application of an established non partial wave approach to few-body atomic bound states has been taken. The necessity of using this non partial wave approach comes from this fact that in few-body atomic calculations one needs a large number of partial wave components, which is caused by very strong short range repulsion of interatomic interactions, to reach proper converged results.
Instead of using standard partial wave representation which leads to tedious and cumbersome numerical procedure we intend to extend a non partial wave approach which has been successfully applied to few-body nuclear systems. In the first step toward this goal the matrix elements of transition amplitudes which appear explicitly in the few-body calculations have been calculated directly as function of two-body Jacobi momentum vectors. The calculated matrix elements can be entered in kernel of three-dimensional Faddeev-Yakubovsky integral equations to study the $^4$He trimer and tetramer ground and exited states. The numerical calculations for these bound states are currently underway.

\section*{Acknowledgments}
We would like to thank the Brazilian agencies FAPESP and CNPq for partial support.


\begin{thebibliography}{99}

\bibitem{Nielsen-JPB3} E. Nielsen, D. V. Fedorov, and A. S. Jensen, J. Phys. {\bf B 31}, 4085 (1998).

\bibitem{Roudnev-CPL328} V. Roudnev and S. L. Yakovlev, Chem. Phys. Lett. {\bf 328}, 97 (2000).

\bibitem{Kolganova-JPB31} E. A. Kolganova, A. K. Motovilov, S. A. Sofianos, J. Phys. {\bf B 31}, 1279 (1998).

\bibitem{Schellingerhout-PRA40} N. W. Schellingerhout, L. P. Kok, and G. D. Bosveld, Phys. Rev. {\bf A 40}, 5568 (1989).

\bibitem{Schellingerhout-FBS11} L. P. Kok, N. W. Schellingerhout, Few-Body Syst. {\bf 11}, 99 (1991).

\bibitem{Carbonell-FBS15} J. Carbonell, C. Gignoux, and S. P. Merkuriev, Few-Body Syst. {\bf 15}, 15 (1993).

\bibitem{Kostrykin-FBS6} V. V. Kostrykin, A. A. Kvitsinsky, and S. P. Merkuriev, Few-Body Syst. {\bf 6}, 97 (1989).

\bibitem{Roudnev-FBS37} V. A. Roudnev, S. L. Yakovlev, and S. A. Sofianos, Few-Body Syst. {\bf 37}, 179 (2005).

\bibitem{Platter-PRA70} L. Platter, H. W. Hammer, Ulf-G. Meissner, Phys. Rev. {\bf A 70}, 052101 (2004).

\bibitem{Platter-FBS35} L. Platter, H. W. Hammer, Ulf-G. Meissner, Few Body Syst. {\bf 35}, 169 (2004).

\bibitem{Platter-PLB607} L. Platter, H. W. Hammer, Ulf-G. Meissner, Phys. Lett. {\bf B 607}, 254 (2005).

\bibitem{Elster-FBS24}
Ch. Elster, J. H. Thomas, and W. Gl\"{o}eckle, Few Body Syst. {\bf 24}, 55 (1998).



\bibitem{Aziz_JCP79}
R. A. Aziz, V. P. S. Nain, {\it et al.}, J. Chem. Phys. {\bf 79}, 4330 (1979).

\bibitem{Aziz_MP61}
R. A. Aziz, F. R. W. McCourt, and C. C. K. Wong, Mol. Phys. {\bf 61}, 1487 (1987).

\bibitem{Aziz_JCP94}
R. A. Aziz and M. J. Slaman, J. Chem. Phys. {\bf 94}, 8047 (1991).

\bibitem{Tang_PRL74}
K. T. Tang, J. P. Toennies, and C. L. Yiu, Phys. Rev. Lett. {\bf 74}, 1546 (1995).

\bibitem{Kolganova-PPN40}
E. A. Kolganova, A. K. Motovilov, and W. Sandhas, Phys. Part. Nucl. {\bf 40}, 2 (2009).

\bibitem{Grisenti-PRL85}
R. E. Grisenti \textit{et al.}, Phys. Rev. Lett. {\bf 85}, 2284 (2000).

\bibitem{Lapack}
the routines called DGESV from {http://netlib.org/lapack/double/}

\bibitem{Elster-FBS27}
Ch. Elster, W. Schadow, A. Nogga, W. Gl\"{o}ckle, Few Body Syst. {\bf 27}, 83 (1999).

\bibitem{Liu-FBS33}
H. Liu, Ch. Elster, W. Gl\"{o}ckle, Few Body Syst. {\bf 33}, 241 (2003).

\bibitem{Hadizadeh-FBS40}
M. R. Hadizadeh and S. Bayegan, Few Body Syst. {\bf 40}, 171 (2007).

\bibitem{Hadizadeh-EPJA36}
M. R. Hadizadeh and S. Bayegan, Eur. Phys. J. A {\bf 36}, 201 (2008).


\end{thebibliography}
\end{document}